\documentclass[a4paper,11pt]{article}
\pdfoutput=1 % if your are submitting a pdflatex (i.e. if you have
\usepackage{jcappub} % for details on the use of the package, please
\usepackage[T1]{fontenc} % if needed
\usepackage{graphicx}
\usepackage{mathtools}
\usepackage{mathrsfs}
\usepackage{epsfig,amssymb,amsmath,color}
\usepackage[colorlinks=true,urlcolor=blue,anchorcolor=blue,citecolor=blue,filecolor=blue,linkcolor=blue,menucolor=blue,pagecolor=blue,linktocpage=true,pdfproducer=medialab,pdfa=true]{hyperref}
\numberwithin{equation}{section}

\usepackage[dvipsnames]{xcolor}

\usepackage{bm,physics}
\usepackage{extarrows}
\usepackage{pgfplots}
\usepackage{graphicx}
\usepackage{tikz}
\usepackage{hyperref}

\title{Smooth coarse-graining and colored noise dynamics in stochastic inflation}

\author[a,1]{Rafid Mahbub,\note{Corresponding author.}}
\author[a]{Aritra De}

\affiliation[a]{School of Physics \& Astronomy\\University of Minnesota, Minneapolis, MN 55455, USA}

\emailAdd{mahbu004@umn.edu}
\emailAdd{aritrade@umn.edu}

\abstract{We consider stochastic inflation coarse-grained using a general class of exponential filters. Such a coarse-graining prescription gives rise to inflaton-Langevin equations sourced by colored noise that is correlated in $e$-fold time. The dynamics are studied first in slow-roll for simple potentials using first-order perturbative, semi-analytical calculations which are later compared to numerical simulations. Subsequent calculations are performed using an exponentially correlated noise which appears as a leading order correction to the full slow-roll noise correlation functions of the type $\big\langle \xi(N)\xi(N') \big\rangle_{(n)}\sim\left( \cosh\left[ n(N-N')+1 \right] \right)^{-1}$. We find that the power spectrum of curvature perturbations $\mathcal{P}_{\zeta}$ is suppressed at small $e$-folds, with the suppression controlled by $n$. Furthermore, we use the leading order, exponentially correlated noise and perform a first passage time analysis to compute the statistics of the stochastic $e$-fold distribution $\mathcal{N}$ and derive an approximate expression for the mean number of $e$-folds $\big\langle \mathcal{N} \big\rangle$. Comparing analytical results with numerical simulations of the inflaton dynamics, we show that the leading order noise correlation function can be used as a very good approximation of the exact noise, the latter being more difficult to simulate.}

\arxivnumber{2204.03859}

\begin{document}

%%%%%%%%%%%%%%%%%%%%%%%%%%%%%%%%%%%%%%%%%%%%%%%%%%%%%%%%%%%%%%%%%%%%%%%%%%%%%%
\maketitle
\flushbottom

\section{Introduction}
It is now well accepted that a period of exponential expansion during the very early Universe, known as cosmic inflation, took place that helped solve the cosmological puzzles that plagued the standard Big Bang model
\cite{Starobinsky:1980te,Sato:1980yn,Guth:1980zm,Linde:1981mu,Albrecht:1982wi,Linde:1983gd}. Apart from solving these cosmological puzzles, the real power of inflation was realized when it was discovered that the quantum vacuum fluctuations of the inflaton field get stretched to cosmological scales and give rise to fluctuations in the matter-energy density of the post-inflationary Universe. These subsequently give rise to structure through gravitational instabilities and the temperature anisotropies ($\delta T(\hat{\bm{n}})/T$) detected in the Cosmic Microwave Background (CMB) radiation \cite{Starobinsky:1979ty,Mukhanov:1981xt,Starobinsky:1982ee,Guth:1982ec}.\\
\indent The standard practice in studying inflationary dynamics is to split the inflaton field into a homogeneous field $\bar{\phi}(t)$, describing the classical background evolution, and small quantum fluctuations $\delta\phi(t,\bm{x})$ which give rise to structure. In this standard prescription, the equation of motion is expanded to leading order in $\delta\phi$ where it is assumed that $\bar{\phi}$ satisfies the inflaton Klein-Gordon equation. This allows one to separate out the dynamics of the quantum fluctuations, which is usually performed by evolving the Fourier modes of the perturbations from subhorizon Bunch-Davies vacuum state up to the superhorizon limit. However, one cannot ignore the role these quantum fluctuations play in modifying the classical evolution of the background inflaton dynamics. The effect of these `quantum kicks' (or backreaction) on the inflaton evolution has come to be known as stochastic inflation \cite{Starobinsky:1986fx,Nambu:1987ef,NAMBU1989240,Matacz:1996gk}. Stochastic inflation is an effective field theory that allows us to study a coarse-grained description of inflation by setting an appropriate classicalization length scale and integrating out degrees of freedom which are smaller. This classicalization of the quantum fluctuations, at the appropriate superhorizon limit, permits a quantum-to-classical transition which effectively enables the quantum kicks to be modeled as Gaussian random noise. Hence, such a procedure provides a description of the evolution of the inflaton through a stochastic differential equation (SDE), similar to how Brownian motion is modeled, that we call the `inflaton-Langevin equation'. Results from stochastic inflation have also been shown to be in agreement with field theoretic calculations \cite{Starobinsky:1994bd,Tsamis:2005hd,Garbrecht:2013coa,Onemli:2015pma}. The effects of quantum diffusion is made all the more apparent when the inflaton reaches a flat portion of the potential where a very small $\dot{\phi}$ allows quantum diffusion to dominate over classical drift. This has led to stochastic inflation gaining a lot of popularity in recent times since formation of primordial black holes (PBHs) can be explained using it. In fact, inflaton potentials containing a non-attractor region like ultra slow-roll can experience amplified quantum diffusion. The increased activity of quantum diffusion leads to an additional enhancement in the curvature power spectrum or a modification of the tail of the distribution of curvature perturbations, which leads to copious PBH formation \cite{Biagetti:2018pjj,Ezquiaga:2018gbw,Ezquiaga:2019ftu,Vennin:2020kng,Figueroa:2020jkf,Pattison:2021oen}.\footnote{We would be remiss not to mention that the situation with a plateau is more complicated to analyze, introducing a fine-tuning issue. In general, if the inflaton does not possess enough kinetic energy to cross the plateau, free diffusion along its extent will lead to PBH overproduction \cite{Rigopoulos:2021nhv}. We thank Ashley Wilkins for pointing this out.}\\
\indent Since stochastic inflation relies on a separation of length scales, there is some ambiguity as to how such coarse-graining is carried out. A smoothing (or filter) function, $W(k/\sigma aH)$, is often employed to delineate between sub and super Hubble regimes where $\sigma\ll 1$ is a parameter that sets the coarse-graining scale. The exact nature of $W$ is not known and a sharp cut-off in Fourier space, accomplished using a Heaviside step function, has been the widely used choice for its mathematical simplicity -- if not for anything else. Such a smoothing function simplifies analytical calculations and describes the inflaton-Langevin equations sourced by Gaussian white noise. White noise is a form of random sourcing where the perturbations are uncorrelated in time (memory-less) and can be described as derivative of Brownian motion (in a formal sense). However, such a simplified smoothing procedure is not very physical and has been known for quite a while that something other than a sharp cut-off produces colored noise (noise with memory). Although more realistic, the latter poses challenges from a modeling point-of-view, both in slow-roll and beyond, since it does not have simple a discretization procedure unlike Brownian motion. As a result, the colored noise needs to be sampled from another stochastic process driven by white noise such that the colored noise generated captures the required statistical properties (usually some predefined two-point correlation function). One of the primary sources of difficulty lies in the dearth of known stochastic processes that can well replicate the desired two-point correlation function of the noise.\\
\indent In this work, $e$-fold time starts from when we start to integrate the background coarse- grained inflaton field. When we consider early $e$-fold time, we are talking about early in that particular inflaton evolution. If we consider the standard inflationary paradigm, these modes correspond to the ones which re-enter the horizon late as the large astrophysical scales. 

\indent In this paper, we study stochastic inflationary dynamics sourced by colored noise. We derive the two-point correlation function of the noise arising from a general class of exponential coarse-graining of the sub Hubble modes. We will show that such a choice of coarse-graining produces correlated noise of the form $\big\langle \xi_{\phi}(N)\xi_{\phi}(N') \big\rangle_{(n)}\propto\left( \cosh\left[ n(N-N') \right]+1 \right)^{-1}$ in Sec. \ref{sec:colored_noise_derivation}. In Sec. \ref{sec:perturb_analysis}, the stochastic inflationary dynamics is studied analytically using the perturbative techniques described in Refs. \cite{Kunze:2006tu,Finelli:2008zg,Finelli:2010sh} for the quadratic and Starobinsky potentials. We compute the curvature power spectrum $\mathcal{P}_{\zeta}$ for the exact noise correlation function along with their leading order expansions. We show that the presence of colored noise modifies the power spectrum at very early $e$-folds as a suppression -- with the $e$-fold length of the suppression dependent on $n$. In Sec. \ref{sec:N_stats}, we formulate a first passage time analysis by deriving an approximate Fokker-Planck equation. This is done by considering the leading order contribution to the colored noise in the form of a decaying exponential. In Sec. \ref{sec:colored_noise_numerical}, the stochastic inflationary dynamics is studied numerically in the presence of colored noise. The colored noise is generated with an auxiliary SDE. In this work we use the natural system of units where $c=\hbar=1$. The reduced Planck mass, $M^{2}_{\text{pl}}=(8\pi G)^{-1}$, is kept in analytical expressions and only set to unity in numerical calculations. Throughout the text, $(...)'$ will be used to refer to a number of things, but the notational significance of the primes will always be followed by appropriate explanations as to what they represent.

\section{Stochastic inflation: a recapitulation}
In describing stochastic inflation, we first recall that the inflaton equations in phase space are given by the following coupled differential equations
\begin{align}
\frac{\dd\phi}{\dd N}&=\pi_{\phi}\label{eq:inflaton1}\\
\frac{\dd\pi_{\phi}}{\dd N}&=-(3-\epsilon_{1})\pi_{\phi}-\frac{\partial_{\phi}V}{H^2}\label{eq:inflaton2}
\end{align}
Here $\pi_{\phi}$ is a field canonically conjugate to $\phi$. The parameter, $\epsilon_{1}=\frac{1}{2M_{\text{pl}}^2}(\frac{\dd\phi}{\dd N})^2$, is the first Hubble flow parameter and
\begin{equation}
H^2=\frac{V}{M_{\text{pl}}^2 (3 - \epsilon_{1})}
\end{equation}
The general prescription in studying inflationary dynamics is to split the $\phi$ and $\pi_{\phi}$ fields into perturbative quantum corrections and treat them separately. This method produces the inflaton Klein-Gordon equation and the Mukhanov-Sasaki equation for the mode evolution. However, we cannot ignore the effects of the quantum fluctuations on the background evolution.\\
\indent In stochastic inflation, we start with the usual decomposition of the inflaton field (and its conjugate) into classical and quantum parts\footnote{Often authors may also denote this split in the following way $$ \phi(N,\bm{x})=\phi_{\text{IR}}(N)+\phi_{\text{UV}}(N,\bm{x}) $$ where UV and IR represent the short and long wavelength modes of the inflaton field in reference to the UV and IR regimes of the electromagnetic spectrum.}
\begin{align}
\phi(N,\bm{x})&=\bar{\phi}(N)+\delta\hat{\phi}(N,\bm{x})\\
\pi_{\phi}(N,\bm{x})&=\bar{\pi}_{\phi}(N)+\delta\hat{\pi}_{\phi}(N,\bm{x})
\end{align}
where $\delta\hat{\phi}(N,\bm{x})$ and $\delta\hat{\pi}_{\phi}(N,\bm{x})$ are sub Hubble quantum fluctuations. These can be expanded in a Fourier basis \cite{Ezquiaga:2018gbw,Pattison:2019hef}
\begin{align}
\delta\hat{\phi}(N,\bm{x})&=\int_{k>0}\frac{\dd^{3}k}{(2\pi)^{3/2}}\widetilde{W}\left( \frac{k}{k_{\sigma}} \right) e^{i\bm{k}\cdot\bm{x}}\delta\phi_{\bm{k}}(N)\hat{a}_{\bm{k}} +\text{h.c.}\label{eq:sub_field1}\\
\delta\hat{\pi}_{\phi}(N,\bm{x})&=\int_{k>0}\frac{\dd^{3}k}{(2\pi)^{3/2}}\widetilde{W}\left( \frac{k}{k_{\sigma}} \right) e^{i\bm{k}\cdot\bm{x}}\delta\pi_{\bm{k}}(N)\hat{a}_{\bm{k}} +\text{h.c.}\label{eq:sub_field2}
\end{align}
The function $\widetilde{W}$ is a window function (in Fourier space) that selects out modes that are $k\gg\sigma aH$ ($k_{\sigma}=\sigma aH$) for $\sigma\ll 1$. We see that the parameter $\sigma$ is used to set the coarse-graining scale in that $k\ll\sigma aH$ are the super Hubble regions which contain the coarse-grained fields $\bar{\phi}$ and $\bar{\pi}_{\phi}$. Window function should have the property that $\widetilde{W}(x)\simeq 1$ for $x\gg 1$. Alternatively, one can equally well reproduce the action of a window function by using a variable lower limit in the integral since the window function effectively removes certain Fourier modes from the integration. Inserting Eq. \eqref{eq:sub_field1} and \eqref{eq:sub_field2} into Eq. \eqref{eq:inflaton1} and \eqref{eq:inflaton2}, we obtain the coarse-grained inflaton equations
\begin{align}
\frac{\dd\bar{\phi}}{\dd N}&=\bar{\pi}_{\phi}+\hat{\xi}_{\phi}\label{eq:cg1}\\
\frac{\dd\bar{\pi}_{\phi}}{\dd N}&=-(3-\epsilon_{1})\bar{\pi}_{\phi}-\frac{\partial_{\phi}V(\bar{\phi})}{H^2(\bar{\phi},\bar{\pi}_\phi)}+\hat{\xi}_{\pi}\label{eq:cg2}
\end{align}
The coarse-grained inflaton equations are now sourced by $\hat{\xi}_{\phi}$ and $\hat{\xi}_{\phi}$, which are given by
\begin{align}
\hat{\xi}_{\phi}&=-\int\frac{\dd^{3}k}{(2\pi)^{3/2}}\frac{\partial}{\partial N}\widetilde{W}\left( \frac{k}{k_{\sigma}} \right)e^{i\bm{k}\cdot\bm{x}}\phi_{\bm{k}_{\sigma}}(N)\hat{a}_{\bm{k}}+\text{h.c.}\\
\hat{\xi}_{\pi}&=-\int\frac{\dd^{3}k}{(2\pi)^{3/2}}\frac{\partial}{\partial N}\widetilde{W}\left( \frac{k}{k_{\sigma}} \right)e^{i\bm{k}\cdot\bm{x}}\pi_{\bm{k}_{\sigma}}(N)\hat{a}_{\bm{k}}+\text{h.c.}
\end{align}
The two-point correlation functions of these operators can be evaluated as $\bra{0}\xi_{f}(N)\xi_{g}(N')\ket{0}$ for a given window function with $f$ and $g$ as placeholder variables that can represent $\phi$ and or $\pi_{\phi}$. The simplest window function is a sharp cut-off with a Heaviside step function. If one considers the quantum-to-classical transition that cosmological perturbations undergo at large scales \cite{Polarski:1995jg,Lesgourgues:1996jc,Kiefer:2008ku,Sudarsky:2009za}, the quantum operators can be replaced by stochastic processes defined by noise two-point correlation functions $\big\langle \xi_{f}(N)\xi_{g}(N') \big\rangle$. This can be understood as foollows. At $k_\sigma$, $\delta\phi_{\bm{k}_\sigma}$ reach their superhorizon values. The $\hat{\xi}$ operators are effectively described by one operator $\hat{b}_{\bm{k}}=\hat{a}_{\bm{k}}\pm \hat{a}^{\dagger}_{-\bm{k}}$ such that the statistics of $b_{\bm{k}}$, $\langle b_{\bm{k}}\rangle=0$ and $\langle b_{\bm{k}} b_{\bm{k}'}\rangle=(2\pi)^3 \delta^{(3)}(\bm{k}+\bm{k}')$, describe a Gaussian random variable. Equations \eqref{eq:cg1} and \eqref{eq:cg2} can now be called `inflaton-Langevin' equations with the name implying that the inflaton evolution is influenced by `random forces'. This is not unlike a particle undergoing Brownian motion, a phenomenon described usually by \textit{Langevin equation}. When $W$ is modeled using a step function, the noise correlations are given by (considering spatial patches are maximally correlated within $k_\sigma$)
\begin{align}
\Big\langle \xi_{\phi}(N)\xi_{\phi}(N') \Big\rangle&=\frac{1}{6\pi^2}\frac{\dd k_{\sigma}^3}{\dd N}|\delta\phi_{k_{\sigma}}(N)|^{2}\delta(N-N')\\
\Big\langle \xi_{\pi}(N)\xi_{\pi}(N') \Big\rangle&=\frac{1}{6\pi^2}\frac{\dd k_{\sigma}^3}{\dd N}|\delta\pi_{k_{\sigma}}(N)|^{2}\delta(N-N')\\
\end{align}
The Dirac $\delta$ functions in the noise two-point functions signify that the noise are `white', defined by it being uncorrelated in $e$-fold time (or even cosmic or conformal time if one were to perform the calculations in such a way). It is, in fact, one of the appealing features of using a sharp cut-off. However, this is rather unphysical and it is expected that these noise terms be correlated at different times and decay with some characteristic time. In the next section, the noise correlation function using a more physical coarse-graining will be derived.

\section{Colored noise from a general class of exponential coarse-graining}\label{sec:colored_noise_derivation}
\subsection{Noise two-point correlation functions}
A sharp cut-off in Fourier space leads to inflaton-Langevin equations sourced by white noise, making them easier to study due to the known properties of the Wiener process. Since white noise processes are Markovian, the SDEs can be discretized and numerically solved in a straightforward fashion (see Ref. \cite{10.1137/S0036144500378302}). However, this sharp cut-off, despite its simplicity, is idealized. In Ref. \cite{Winitzki:1999ve}, a Gaussian filter was considered and modeled as follows\footnote{The Gaussian filter was subsequently used in Refs. \cite{Matarrese:2003ye,Liguori:2004fa} to study its effects on $\mathcal{P}_{\zeta}$ and low-$\ell$  multipoles of the CMB power spectrum.}
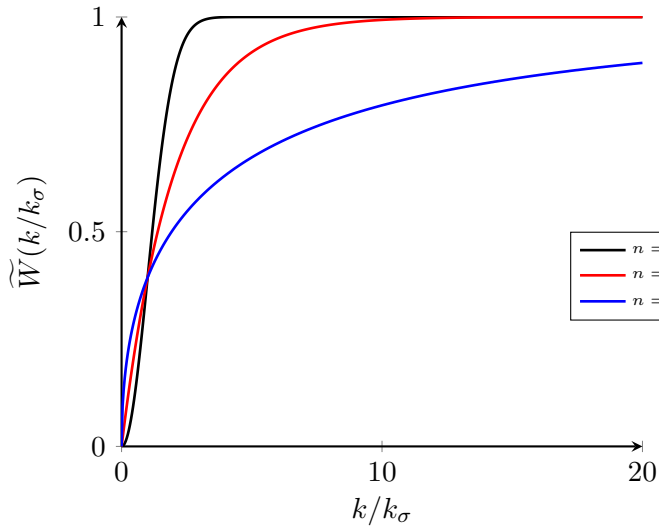
\begin{figure} 
\centering
 \begin{tikzpicture}
     \begin{axis}
      [
        axis lines = left,
        axis line style = thick,
        xlabel = \( k/k_\sigma \),
        ylabel = {\( \widetilde{W}(k/k_\sigma) \)},
        xtick = {0, 10, 20},
        ytick = {0, 0.5, 1.0},
        legend style={at={(1,0.5)},anchor=north,legend cell align=left,font=\tiny}
      ]
      \addplot
      [
        domain=0:20,
        samples=500,
        line width=1.0pt,
        color=black,
      ]
      {1-exp(-0.5*x^2)};
      \addlegendentry{\( n=2 \)}
      
      \addplot
      [
        domain=0:20,
        samples=500,
        line width=1.0pt,
        color=red,
      ]
      {1-exp(-0.5*x)};
      \addlegendentry{\( n=1 \)}
      
      \addplot
      [
        domain=0:20,
        samples=500,
        line width=1.0pt,
        color=blue,
      ]
      {1-exp(-0.5*x^0.5)};
      \addlegendentry{\( n=1/2 \)}
     \end{axis}
 \end{tikzpicture}
 \caption{A graphical illustration of the window function for three different values of $n$ where we are able to see the speed at which $W(x)$ approaches unity for different values of the index.}
 \label{fig:window_function}
 \end{figure}
\begin{equation}\label{eq:gaussian_filter}
\widetilde{W}\left( \frac{k}{k_{\sigma}} \right)=1-e^{-k^{2}/2k_{\sigma}^2}
\end{equation}
However, a Gaussian is not the only option and any function can be chosen as long as the properties of a window function are satisfied. Here we consider a general class of exponential filters of the form
\begin{equation}\label{eq:exp_filter}
\widetilde{W}_{(n)}\left( \frac{k}{k_{\sigma}} \right)=1-\exp\left[ -\frac{1}{2}\left( \frac{k}{k_{\sigma}} \right)^n \right]
\end{equation}
where it reduces to the Gaussian case for $n=2$. This filter possesses the correct properties required for coarse-graining in the stochastic inflationary approach, namely $\widetilde{W}(k/k_{\sigma})\simeq 0$ when $k/k_{\sigma}\ll 1$. In some sense, $n$ controls the `sharpness' of the cut-off with $n\rightarrow\infty$ representing a sharp cut-off, as can be seen in Fig. (\ref{fig:window_function}). Using the definition of the noise term, we may express its correlation function as follows
\begin{align}\label{eq:phi-phi1}
\Big\langle \xi_{\phi}(N_1,\bm{x}_1)\xi_{\phi}(N_2,\bm{x}_2) \Big\rangle_{(n)}&=\int\frac{\dd^{3}k}{(2\pi)^3}\frac{\partial}{\partial N_1}\widetilde{W}_{(n)}\left( \frac{k}{k_{\sigma}} \right)\frac{\partial}{\partial N_2}\widetilde{W}_{(n)}\left( \frac{k}{k_{\sigma}} \right)e^{i\bm{k}\cdot(\bm{x}_1-\bm{x}_2)}\delta\phi_{\bm{k}}(N_1)\delta\phi_{\bm{k}}^{*}(N_2)
\end{align}
Now, it is straightforward to show that
\begin{equation}
\frac{\partial}{\partial N}\widetilde{W}_{(n)}\left( \frac{k}{k_{\sigma}}\right)=\frac{n}{2}\left( \frac{k}{k_{\sigma}} \right)^{n}(\epsilon_{1}-1)e^{-\frac{1}{2}\left( \frac{k}{k_\sigma} \right)^n}
\end{equation}
For simplicity, we evaluate the correlation function at equal spatial points. Then, Eq. \eqref{eq:phi-phi1} can be written as
\begin{align}\label{eq:phi-phi2}
\Big\langle \xi_{\phi}(N_1)\xi_{\phi}(N_2) \Big\rangle_{(n)}&=\int_{0}^{\infty}\frac{\dd k}{2\pi^2}k^{2}\left( \frac{k}{\sigma a_1 H_1} \right)^{n}\left( \frac{k}{\sigma a_2 H_2} \right)^{n}\big(\epsilon_{1}(N_1)-1\big)\big(\epsilon_{1}(N_2)-1\big)\nonumber\\
&\times \exp\left[ -\left(\frac{1}{a_1^n H_1^n}+\frac{1}{a_2^n H_2^n} \right)\frac{k^n}{2\sigma^n} \right]\delta\phi_{\bm{k}}(N_1)\delta\phi^{*}_{\bm{k}}(N_2)
\end{align}
where the subscripts on $a$ and $H$ refer to dependence on either $N_1$ or $N_2$ for notational simplicity. Equation \eqref{eq:phi-phi2} represents the noise correlation function of the $\delta\phi_{\bm{k}}$ modes in its complete form. Unlike the white noise case, where the integral is essentially made obsolete due to the presence of a Dirac $\delta$ function, Eq. \eqref{eq:phi-phi2} is much more complicated. Beyond slow-roll, it would require a numerical integration over all the Fourier modes of interest, which we will not endeavor to do. Moreover, determining the functional form the correlation function will also be challenging and would likely need to be fitted with known functions. However, we can still make some progress under the slow-roll assumption where $\epsilon_{1}\simeq 0$ and $H\simeq\text{constant}$. We know that, in the slow-roll limit, the inflaton fluctuations admit a simple, closed-form superhorizon expression \cite{Riotto:2002yw}
\begin{equation}
|\delta\phi_{\bm{k}}|\simeq\frac{H}{\sqrt{2k^3}}\left( \frac{k}{aH} \right)^{3/2-\nu}
\end{equation}
where $\nu^{2}=\frac{9}{4}-\frac{m^{2}}{H^2}$, with $m$ representing the inflaton mass which is assumed to be much smaller than the Hubble scale. Then, working in this approximation
\begin{align}\label{eq:colored_noise0}
\Big\langle \xi_{\phi}(N_1)\xi_{\phi}(N_2) \Big\rangle_{(n)}&\simeq \frac{n^2}{4}\frac{H^2}{8\sigma^{2n}}\left( a_1 H \right)^{\nu-n-3/2}\left( a_2 H \right)^{\nu-n-3/2}\nonumber\\
&\qquad\qquad\qquad\times\int_{0}^{\infty}\frac{\dd k}{2\pi^2}k^{2+2n-2\nu}\exp\left[ -\left(\frac{1}{a_1^n H_1^n}+\frac{1}{a_2^n H_2^n} \right)\frac{k^n}{2\sigma^n} \right]\nonumber\\
&=\frac{H^2 n}{16\pi^2}\sigma^{3-2\nu}\cdot 2^{\frac{1}{n}(3+2n-2\nu)}\frac{\left( a_1^n+a_2^n \right)^{\frac{1}{n}(2\nu-2n-3)}}{(a_1 a_2)^{\nu-n-3/2}}\Gamma\left( \frac{3+2n-2\nu}{n} \right)
\end{align}
For massless fields in de Sitter space, $\nu=3/2$. Then,
\begin{equation}\label{eq:colored_noise}
\boxed{\Big\langle \xi_{\phi}(N_1)\xi_{\phi}(N_2) \Big\rangle_{(n)}\simeq\frac{H^2 n}{4\pi^2}\frac{(a_1^n +a_2^n)^{-2}}{(a_1 a_2)^{-n}}=\frac{H^2}{4\pi^2}\frac{n}{2}\frac{1}{\cosh[n(N_1-N_2)]+1}}
\end{equation}
Equation \eqref{eq:colored_noise} represents the noise correlation function for a general exponential filter derived using the slow-roll and massless de Sitter assumptions. The fact that it is some function other than a Dirac $\delta$ distribution affirms the fact that it represents noise with finite correlation time. It is interesting to note that, in the massless de Sitter limit, the coarse-graining scale, $\sigma$, drops out. More accurately though, one should expect corrections from $\sigma$ in the noise correlation since $\sigma^{3-2\nu}$ can be expanded out (the corrections will be negligible in the slow-roll case). Working out what these corrections might be, we treat $\nu$ as a small expansion parameter, such that
\begin{align}
\sigma^{3-2\nu}&\simeq\sigma^3 \left( 1-2\nu\ln\sigma \right)\\
\Gamma\left( \frac{3+2n-2\nu}{n} \right)&\simeq \Gamma\left( 2+\frac{3}{n} \right)-\frac{2\nu}{n}\Gamma\left( 2+\frac{3}{n} \right)\psi^{(0)}\left( 2+\frac{3}{n} \right)
\end{align}
where $\psi^{(0)}(z)$ is the zeroth-order polygamma function defined as 
\begin{equation}
\psi^{(0)}(z)=\frac{\dd}{\dd z}\ln\Gamma(z)
\end{equation}
 These approximations can be used to study the effects of the coarse-graining scale $\sigma$ on the dynamics. However, one needs to set how small the mass should be and determine whether or not any notable changes occur. In general, different functional representations of $W(k/k_{\sigma})$ can be used, as long as the properties of the filter functions are satisfied. Another function that can mimic the properties of Eq. \eqref{eq:gaussian_filter} rather well is $\tanh(k/k_{\sigma})$. However, analytical integration of $\big\langle \xi\xi \big\rangle$ in slow-roll might not be straightforward and one would need to resort to approximations. Regardless of the choice, they will all lead to finite correlations in $e$-fold, making a numerical implementation even more nontrivial. As we shall see later, SDEs with colored noise are solved by first obtaining an auxiliary SDE that is driven by white noise. The stochastic process which is a solution to the auxiliary SDE must have a two-point correlation function that resembles the colored noise correlation that we are trying to sample. In particular, we focus on three cases
 \begin{equation}
 \Big\langle \xi_\phi(N_1)\xi_\phi(N_2) \Big\rangle_{(n)}=\frac{H^2}{4\pi^2}
    \begin{dcases}
        \frac{1}{2}\frac{1}{2\left( \cosh\left[ \frac{1}{2}(N_1-N_2) \right]+1 \right)} & n=\frac{1}{2}\\
        \frac{1}{2\left( \cosh\left[ (N_1-N_2) \right]+1 \right)} & n=1 \\
        \frac{1}{2\cosh^2(N_1-N_2)} & n=2 
    \end{dcases}
 \end{equation}
 As we shall see later, these different fall-off behaviors of these noise correlation functions will have noticeable effects on the curvature power spectrum in the small $e$-fold range, or in other words, early times in inflation. It is also easy to deduce that, in the limit $n\rightarrow\infty$, this class of noise correlation functions approach the Dirac $\delta$ function of white noise such that the dynamics of a sharp cut-off can be recovered. Since we solely focus on the noise correlation function of the field variable, the subscript $\phi$ will be dropped in the future.

 \subsection{Not all smooth window functions are created equally --- large $r$ behavior}
\begin{figure}
     \centering
     \includegraphics{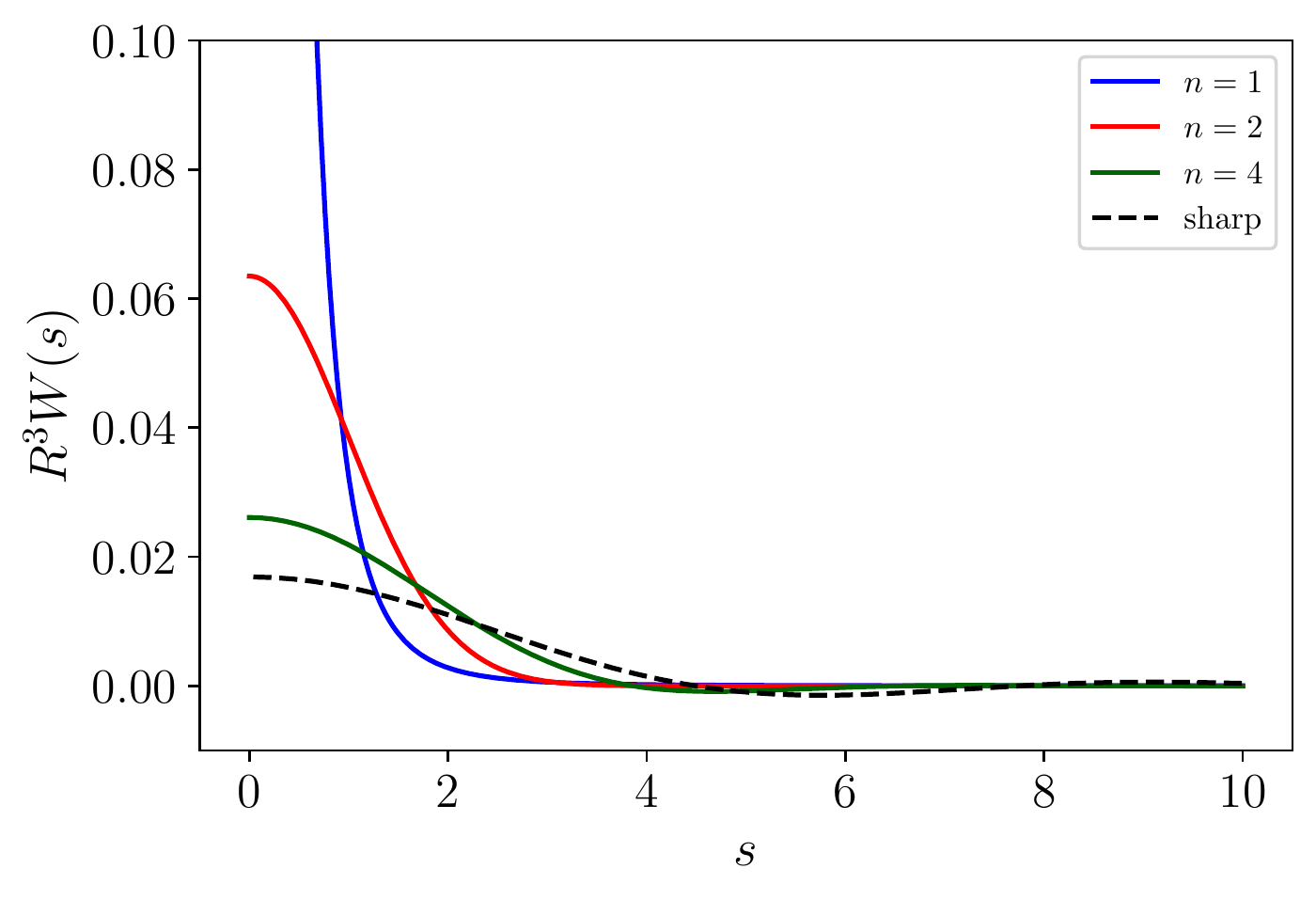}
     \caption{The real space window functions for $n=1,2$ and $4$, compared to the one obtained from the sharp cut-off in momentum space (black dashed).}
     \label{fig:window_real_space}
 \end{figure}

 Here we compare the window functions from the previous section and study their large $r$ fall-off, as performed in Ref. \cite{Winitzki:1999ve}. For the sake of simplicity, we also make minor adjustments to the notation, only for this subsection, in accordance to \cite{Winitzki:1999ve}. There, the authors pointed out the particularly slow decay of $\big\langle \xi(N,\bm{x})\xi(N',\bm{x}') \big\rangle$ for large $|\bm{x}-\bm{x}'|$ for the white noise arising from a sharp cut-off in momentum space.\\
 \indent For a sharp cut-off, we have
 \begin{equation}
     \widetilde{W}_{\text{sharp}}\left( \frac{k}{k_\sigma} \right)=\Theta\left( 1-\frac{k}{k_\sigma} \right)\equiv \Theta\left( 1-kR_\sigma \right)
 \end{equation}
 In real space, this turns out to be
 \begin{equation}
     W_{\text{sharp}}(s)=\frac{\sin s-s\cos s}{2\pi^2 s^3}\;\;\;\;\;\text{where} \quad s=\frac{|\bm{x}|}{R_\sigma}
 \end{equation}
 It has been shown that well-motivated smooth window functions which produce the predictions independent of the choice of smoothing should behave as $W(s)\sim s^{-6}$ for $s\rightarrow\infty$. Furthermore, a corollary condition that was established reads
 \begin{equation}\label{eq:condition}
     \boxed{\frac{\dd^2}{\dd u^2}\widetilde{W}(u)\bigg\lvert_{u=0}=-\frac{4\pi}{3}\int_{0}^{\infty} s^4 W\left( s \right)\dd s<0}
 \end{equation}
 Here $u=kR_\sigma$. It is easy to check that the sharp momentum space cut-off is ill-behaved in real space, large distance asymptotics, varying as $W_{\text{sharp}}(s)\sim s^{-2}$. As a result, Eq. \eqref{eq:condition} is highly divergent. We now check the large distance properties of the general class of exponential filters, focusing on the cases $n=1,2$ and $4$, with a generalized analysis on the asymptotics provided in Appendix \ref{sec:asymptotics} for even $n$. We first define the inverse Fourier transform that maps the $k$-space window function to real space
 \begin{equation}
     W_{(n)}\left( \frac{r}{R_\sigma} \right)=\int\frac{\dd^3 k}{(2\pi)^3}e^{i\bm{k}\cdot\bm{x}}\widetilde{W}_{(n)}(kR_\sigma)
 \end{equation}
 This yields the following
 \begin{equation}\label{eq:real_space_windows}
 R_\sigma^3 W_{(n)}\left( s \right)=
     \begin{dcases}
        \frac{8}{\pi^2}\frac{1}{\left(1+4 s^2\right)^2}&n=1\\
        \frac{1}{\left( 2\pi \right)^{3/2}}e^{-s^2/2}&n=2\\
     \end{dcases}
 \end{equation}
 We have not written down an explicit expression for $W_{(n=4)}$. However, we will make use of the appropriate asymptotic expression from Appendix \ref{sec:asymptotics}. The real space window functions are plotted in Fig (\ref{fig:window_real_space}) and compared to the one obtained from a sharp $k$-space cut-off. From the figure, although it is difficult to ascertain the large $s$ behavior, one can see that the $n=4$ window becomes slightly negative at certain points along the real $s$ line. However, these variations are not as prominent as $W_\text{sharp}$.\\
 \indent Turning our attention to Eq. \eqref{eq:real_space_windows}, we can try to understand the asymptotics of the different real space window functions. For $n=1$, $W_{(n=1)}\sim s^{-4}$, which is not ideal but, nevertheless, fares much better than $W_\text{sharp}$. The asymptotics of $W_{(n=4)}$ can be extracted using Eq. \eqref{eq:asymptotic_general}, from which we obtain
 \begin{equation}
     R_\sigma^3 W_{(n=4)}\sim -\frac{1}{4\pi s}\sqrt{\frac{\pi}{3}}\exp\left(-\frac{3}{2^{10/3}}s^{4/3}\right)\sin\left( \frac{3^{3/2}}{2^{10/3}}s^{4/3}+\frac{\pi}{6} \right)
 \end{equation}
 \\
Finally, we evaluate the different smooth window functions, against $W_\text{sharp}$, using the criterion in Eq. \eqref{eq:condition}. From the onset, we can say that only $n=2$ completely satisfies the two necessary conditions -- in that its decay is faster than $s^{-6}$ and 
\begin{equation}
    \frac{\dd^2}{\dd u^2}\widetilde{W}_{(n=2)}(u)\bigg\lvert_{u=0}=-\frac{1}{R_\sigma^3}<0\;\;\;\;(\text{bounded})
\end{equation}
With $n=4$, the dominant contribution to the asymptotics arises from the $e^{-s^{4/3}}$ term. We can show that the criterion in Eq. \eqref{eq:condition} is satisfied
\begin{equation}
    \frac{\dd^2}{\dd u^2}\widetilde{W}_{(n=4)}(u)\bigg\lvert_{u=0}=-\frac{32}{27}\sqrt{\frac{\pi}{3}}\frac{1}{R^3_\sigma}<0\;\;\;\;(\text{bounded})
\end{equation}
The case with $n=1$ is tricky since one can show that is grows unbounded
\begin{equation}
    \frac{\dd^2}{\dd u^2}\widetilde{W}_{(n=1)}(u)\bigg\lvert_{u=0}\sim -\frac{2}{3\pi}s\;\;\;\;(\text{for large }s)
\end{equation}
In one sense it satisfies the condition in Eq. \eqref{eq:condition} in that it is always negative. However, it is not finite itself and always grows along the negative line as $\mathcal{O}(10^{-1})s$ for large $s$. Nevertheless, they all perform better than the sharp cut-off for which the large $r$ asymptotic behavior is
\begin{equation}
    \frac{\dd^2}{\dd u^2}\widetilde{W}_{\text{sharp}}(u)\bigg\lvert_{u=0}\sim\frac{2}{3\pi}s^2\;\;\;\;(\text{for large }s)
\end{equation}
Finally, we check what happens when $n\rightarrow\infty$. One can prove that $\lim_{n\rightarrow\infty}\Re{f(z_{0,n})}=0$ and $\lim_{n\rightarrow\infty}\Im{f(z_{0,n})}=-2$. As a result, $W_{(n)}$ no longer possesses an exponential suppression such that we end up with
\begin{equation}
    \lim_{n\rightarrow\infty}W_{(n)}(s)\sim s^{-3/2}\sin s
\end{equation}
This shows that the characteristics of the window function approach that of $W_\text{sharp}$ for increasing values of $n$. This is a reasonable statement and very easy to visualize as well. 

\section{Stochastic inflationary correlation functions using perturbative expansion}\label{sec:perturb_analysis}
\subsection{Slow-roll derivation of $\langle \delta\phi^{(1)\;2} \rangle$}
For simple inflation models, under the assumption of slow-roll, stochastic inflationary correlation functions can be (semi) analytically studied by perturbing the inflaton around a background and studying them order-by-order \cite{Finelli:2008zg,Finelli:2010sh,Kunze:2006tu}. In slow-roll, the inflaton-Langevin equation is given by
\begin{equation}\label{eq:langevin}
\frac{\dd\phi}{\dd N}=-2M_{\text{pl}}^{2}\frac{H'}{H}+\frac{H}{2\pi}\circ_{\alpha} \xi
\end{equation} 
where, the $(...)'$ refers to derivatives with respect to the field variable, $\phi$ (As a result, to avoid confusion, different $e$-folds will be labelled using numbered subscripts instead of primes) and $\circ_\alpha$ referring to the discretization ambiguity of stochastic integrals where $\alpha\in[0,1]$. For this work, we choose $\alpha=0$ which corresponds to the It\^{o} interpretation. The $\xi(N)$ is a (normalized) noise term the statistics of which is kept unspecified for now. We also ignore the dynamics of $\pi_\phi$, which is negligible in slow-roll. In other words, we consider for the moment $\big\langle \xi(N_{1})\xi(N_{2}) \big\rangle\sim \delta(N_{1}-N_{2})$. Stochastic effects can be studied through the inclusion of perturbations as follows.
\begin{equation}
\phi=\phi_{\text{cl}}+\sum_{n}\epsilon_{n}\delta\phi^{(n)}
\end{equation}
where $\phi_{\text{cl}}$ is the background inflaton field that satisfies Eq. \eqref{eq:langevin} without the noise term. Here we consider an expansion of the first order, for which it can be show that
\begin{align}
\frac{H'}{H}&\simeq\frac{H'(\phi_{\text{cl}})}{H(\phi_{\text{cl}})}+\left( -\frac{H'^{2}(\phi_{\text{cl}})}{H^{2}(\phi_{\text{cl}})} +\frac{H''(\phi_{\text{cl}})}{H(\phi_{\text{cl}})} \right)\delta\phi^{(1)}\nonumber\\
&=\frac{H'}{H}+\left( \frac{H'}{H} \right)'\delta\phi^{(1)}
\end{align}
Here we do not consider $\delta\phi^{(2)}$ terms, which become important for studies in primordial non-Gaussianities. Using this expansion, one finds that the first order perturbations satisfy the following differential equation
\begin{equation}\label{eq:langevin2}
\frac{\dd\delta\phi^{(1)}}{\dd N}+2M_{\text{pl}}^{2}\left( \frac{H'}{H} \right)'\delta\phi^{(1)}=\frac{H}{2\pi}\xi
\end{equation}
Due to the perturbative expansion, the Hubble parameter and its derivatives, $H$ and $H'$, are all functions of $\phi_{\text{cl}}$. The formal solution to Eq. \eqref{eq:langevin2} can be computed in a straightforward manner

\begin{align}
\delta\phi^{(1)}=\exp\left[-2M_{\text{pl}}^{2}\int^{N}\left( \frac{H'}{H} \right)'\dd N_{1}\right]\int^{N}\;\frac{H}{2\pi}\xi(N_{1})\exp\left[2M_{\text{pl}}^{2}\int^{N_{1}}\left( \frac{H'}{H} \right)'\dd N_{2}\right]\dd N_{1}
\end{align}
Since $\delta\phi^{(1)}$ is a function of $\xi(N)$, it is a quantity that is dependent on the particular realization of the random process that defines the noise -- hence a stochastic process itself. As a result, it is more meaningful to study the ensemble average of the fluctuations $\big\langle \delta\phi^{(1)\;2} \big\rangle$. Since it is an average, it is expected to characterize the model's behavior over a large number of observations and can be used in conjunction with numerical simulations. Moreover, the power spectrum of curvature perturbations can be defined as follows (see Sec. \ref{sec:colored_noise_numerical})
\begin{equation}
\mathcal{P}_{\zeta}=\frac{1}{1-\epsilon_{1}}\frac{\dd}{\dd N}\left( \frac{\Big\langle \delta\phi^{(1)\;2}\Big\rangle}{2M_{\text{pl}}^2\epsilon_{1}} \right)
\end{equation}
A differential equation for $\big\langle \delta\phi^{(1)\;2} \big\rangle$ can be obtained by multiplying $\delta\phi^{(1)}$ to Eq. \eqref{eq:langevin2} and taking expectation values on both sides. Then, we get
\begin{equation}\label{eq:langevin3}
\frac{\dd}{\dd N}\Big\langle \delta\phi^{(1)\;2} \Big\rangle+4M_{\text{pl}}^{2}\left( \frac{H'}{H} \right)'\Big\langle \delta\phi^{(1)\;2} \Big\rangle=\frac{H}{\pi}\Big\langle \xi\delta\phi^{(1)} \Big\rangle
\end{equation}
The source term in Eq. \eqref{eq:langevin3}, $\big\langle \xi\delta\phi^{(1)} \big\rangle$ requires a closer examination since it contains the noise term and it can be used to establish differences between white and colored noise sources. The source term can be expressed as follows--

\begin{equation}\label{eq:expec}
\Big\langle \xi\delta\phi^{(1)}\Big\rangle=\exp\left[-2M_{\text{pl}}^{2}\int^{N}\left( \frac{H'}{H} \right)'\dd N_{1}\right]\int^{N}\;\frac{H}{2\pi}\Big\langle \xi(N)\xi(N_{1})\Big\rangle\exp\left[2M_{\text{pl}}^{2}\int^{N_{1}}\left( \frac{H'}{H} \right)'\dd N_{2}\right]\dd N_{1}
\end{equation}
The expression for $\big\langle \xi\delta\phi^{(1)} \big\rangle$ crucially depends on the noise correlation function. We see that for white noise where $\big\langle \xi(N_{1})\xi(N_{2}) \big\rangle=\delta(N_{1}-N_{2})$, this reduces to
\begin{equation}
\Big\langle \xi\delta\phi^{(1)} \Big\rangle_{\text{white-noise}}=\frac{H}{4\pi}
\end{equation}
which is a slowly varying function. However, for a general colored noise process, Eq. \eqref{eq:expec} might not be exactly solvable depending on the complexity of the functional form of the noise correlation. Fortunately, this can be analytically studied for the general class of exponential window functions. In general, we will be interested in two classes of colored noise terms
\begin{align}
\Big\langle \xi(N_{1})\xi(N_{2}) \Big\rangle&\sim\frac{1}{\cosh\left[ n(N_1-N_2) \right] +1}\;\;\;\textsf{(general exponential coarse-graining)}\label{eq:noise1}\\
\Big\langle \xi(N_{1})\xi(N_{2}) \Big\rangle&\sim e^{-2|N_{1}-N_{2}|}\label{eq:noise2}\qquad\qquad\quad\;\;\;\;(\textsf{Ornstein-Uhlenbeck})
\end{align}
Of particular interest is the noise produced by Eq. \eqref{eq:noise2} which is the leading order contribution to Eq. \eqref{eq:noise1} for $n=2$. Also, from a simulation standpoint, the $e^{-2|N_{1}-N_{2}|}$ correlation function can be studied more naturally since such exponentially decaying noise correlation functions are characteristic of the late-time behavior of noise correlations in Ornstein-Uhlenbeck (OU) processes \cite{PhysRev.36.823,gardiner1985handbook}. Using these, $\big\langle \delta\phi^{(1)\;2} \big\rangle$ can be expressed by the following
\begin{equation}\label{eq:<delphi2>}
\Big\langle \delta\phi^{(1)\;2} \Big\rangle=\exp\left[-4M_{\text{pl}}^{2}\int^{N}\left( \frac{H'}{H} \right)'\dd N_{1}\right]\int^{N}\;\frac{H}{\pi}\Big\langle \xi\delta\phi^{(1)}\Big\rangle\exp\left[4M_{\text{pl}}^{2}\int^{N_{1}}\left( \frac{H'}{H} \right)'\dd N_{2}\right]\dd N_{1}
\end{equation}
For simple potentials, Eq. \eqref{eq:<delphi2>} can be rewritten by changing the variables from $N$ to $\phi_{\text{cl}}$ using $\dd N=-H/(2M_{\text{pl}}^{2}H')\dd\phi_{\text{cl}}$.
\begin{equation}
\Big\langle \delta\phi^{(1)\;2} \Big\rangle=-\frac{1}{2\pi M_{\text{pl}}^2}\frac{H'^{2}}{H^2}\int\dd\phi_{\text{cl}}\frac{H^4}{H'^{3}}\Big\langle \xi\delta\phi^{(1)} \Big\rangle
\end{equation}
For white noise processes, the previous equation further simplifies to
\begin{equation}
\Big\langle \delta\phi^{(1)\;2} \Big\rangle=-\frac{1}{8\pi^{2} M_{\text{pl}}^2}\frac{H'^{2}}{H^2}\int\dd\phi_{\text{cl}}\frac{H^5}{H'^{3}}
\end{equation}
In general, however, it may be easier to use the $e$-fold variable since there may not be a simple expression that allows a mapping between $N$ and $\phi_{\text{cl}}$. For realistic potentials, exact expressions for $\big\langle \delta\phi^{(1)\;2} \big\rangle$ and $\mathcal{P}_\zeta$ will be very challenging to derive using the colored noise correlations. As a result, in what follows, we first study the effects of a finite correlation time on a quadratic inflaton potential which can serve as a good prototype for investigating the effects smooth coarse-graining has on inflationary observables.

We have analytically calculated the small $e$-fold behavior of $\mathcal{P}_{\zeta}$ (early times in inflation) with colored noise in the appendix \ref{small_efold}.

\subsection{Chaotic potential}
\subsubsection{White noise}
\begin{figure*}
\centering
\includegraphics{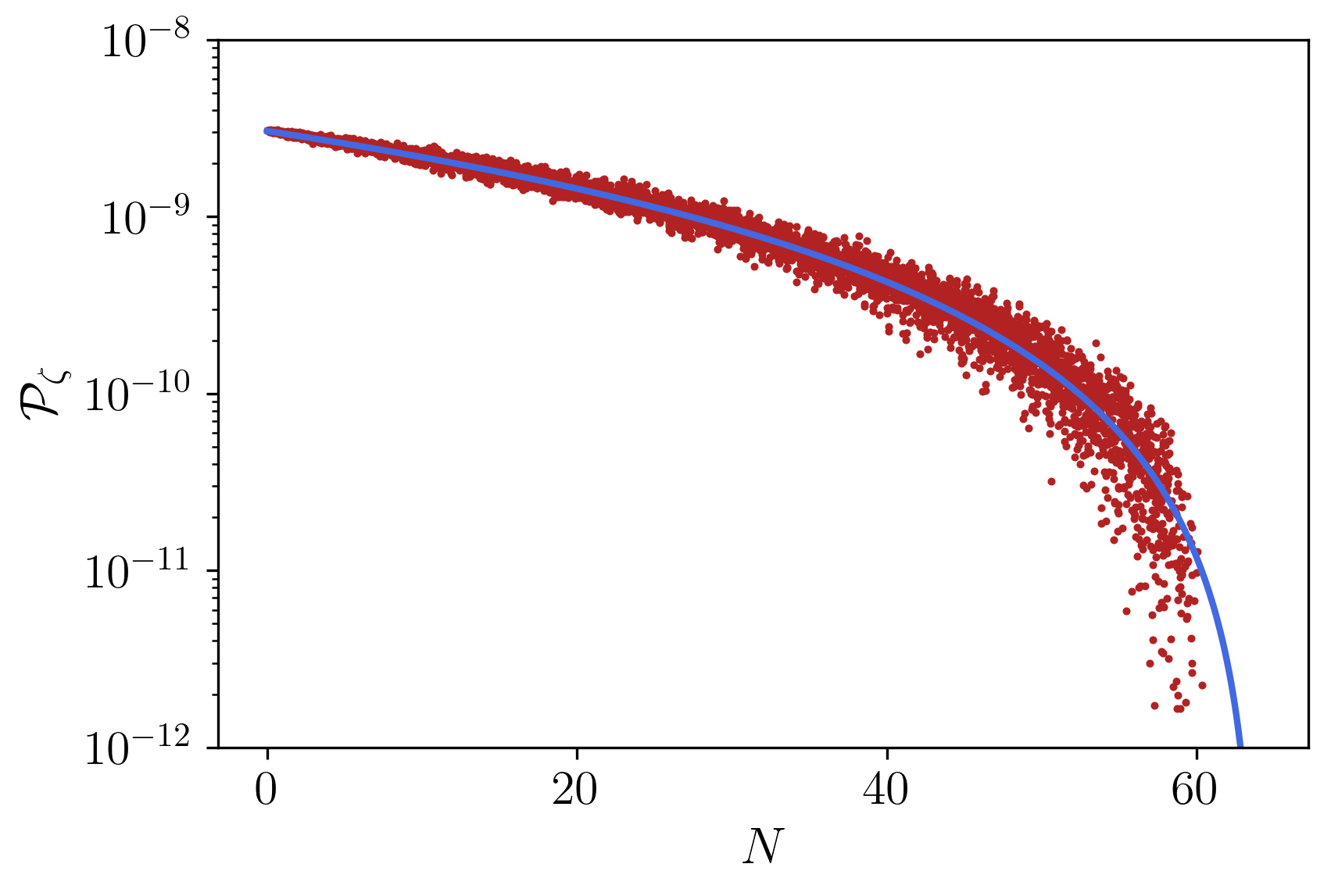}
\caption{Plot showing the curvature power spectrum $\mathcal{P}_{\zeta}$ for the chaotic potential obtained by simulating the inflaton-Langevin equation a large number of times (dotted red) compared to the analytical expression in Eq. \eqref{eq:Pzeta_analytic} (reproduced from \cite{De:2020hdo}).}
\label{fig:Pzeta_white}
\end{figure*}
The analytical expressions can be checked for both white and colored noise processes by considering the simple chaotic inflation potential, $V(\phi)=m^2 \phi^2 /2$ with $m^2 = 4.4\times 10^{-11}M_{\text{pl}}^2$. This particular value of $m$ gives the correct CMB normalization of the power spectrum, $\mathcal{P}_{\zeta}\simeq 2.2\times 10^{-9}$, at $N_{\star}=10$ where the pivot scale ($k_{\star}=0.05\;\text{Mpc}^{-1}$) is set \cite{Akrami:2018odb}. In this section, we drop the `cl' subscripts from $\phi_{\text{cl}}$ for brevity. For our calculations, we consider the inflaton rolling down from an initial field value $\phi_{\text{in}}=16M_{\text{pl}}$ for which inflation lasts for about $64\;e$-folds. Then,
\begin{align}
\Big\langle \delta\phi^{(1)\;2} \Big\rangle&=-\frac{1}{8\pi^{2}M_{\text{pl}}^2}\frac{1}{\phi^2}\int_{\phi_{\text{in}}}^{\phi}\dd\phi'\frac{m^2}{6M_{\text{pl}}^2}\phi'^5\nonumber\\
&=\frac{m^2}{48\pi^2 M_{\text{pl}}^4}\frac{1}{6\phi^2}\left( \phi_{\text{in}}^6 - \phi^6 \right)
\end{align}
To analytically compute $\mathcal{P}_{\zeta}$, we use the fact that in slow-roll $\epsilon_{1}\ll1$. In this approximation
\begin{equation}
\mathcal{P}_{\zeta}\simeq \frac{\dd}{\dd N}\left( \frac{\Big\langle \delta\phi^{(1)\;2} \Big\rangle}{2M_{\text{pl}}^2\epsilon_{1}} \right)
\end{equation}
For the quadratic potential, $\epsilon_{1}=2M_{\text{pl}}^{2}/\phi^2$ and $\mathcal{P}_{\zeta}$ is given by
\begin{equation}\label{eq:Pzeta_analytic}
\mathcal{P}_{\zeta}\simeq\frac{m^2}{96\pi^2 M_{\text{pl}}^6}\phi^4 = \frac{H^2(\phi_{\text{in}})}{8\pi^2 M^2_{\text{pl}}\epsilon_{1}(\phi_{\text{in}})}\left( 1-\frac{4M_{\text{pl}}^{2}N}{\phi^{2}_{\text{in}}} \right)^2
\end{equation}
where $\phi(N)=\sqrt{\phi_{\text{in}}^{2}-4M_{\text{pl}}^{2}N}$ is the solution of the inflaton background evolution for the chaotic potential (in slow-roll). Here $\phi_{\text{in}}$ represents some initial field value at which the inflaton starts rolling down the potential. Consequently, we set $N_{\text{in}}=0$. This calculation reproduces the standard result of $\mathcal{P}_{\zeta}$ in slow-roll. This is shown in Fig. (\ref{fig:Pzeta_white}) where the dotted plot represents $\mathcal{P}_{\zeta}$ calculated by simulating the inflaton-Langevin equations $10^{6}$ times (see Ref. \cite{De:2020hdo}).

\subsubsection{Colored noise: exact and leading order results}
\begin{figure}
\centering
\includegraphics[scale=0.5]{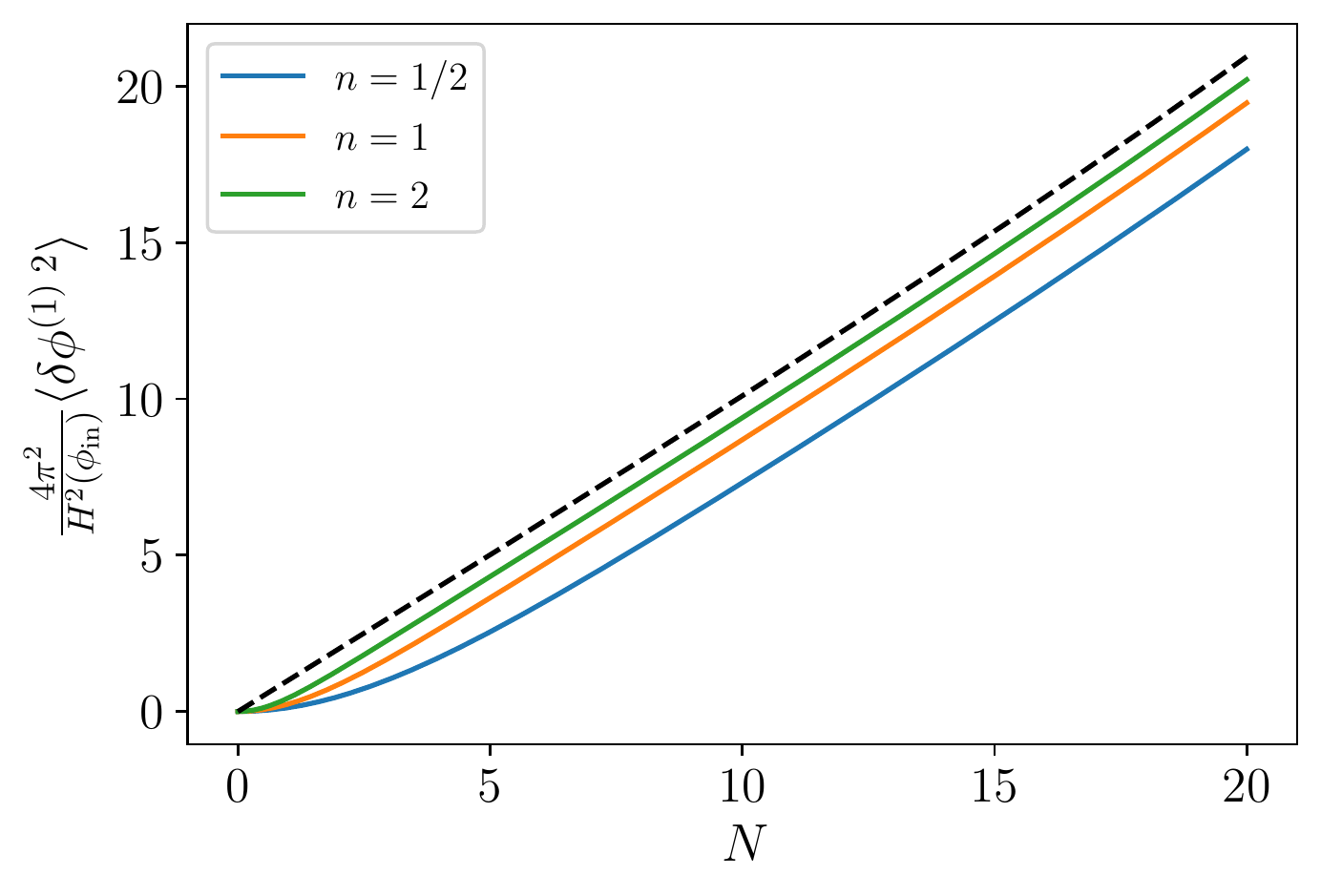}
\includegraphics[scale=0.5]{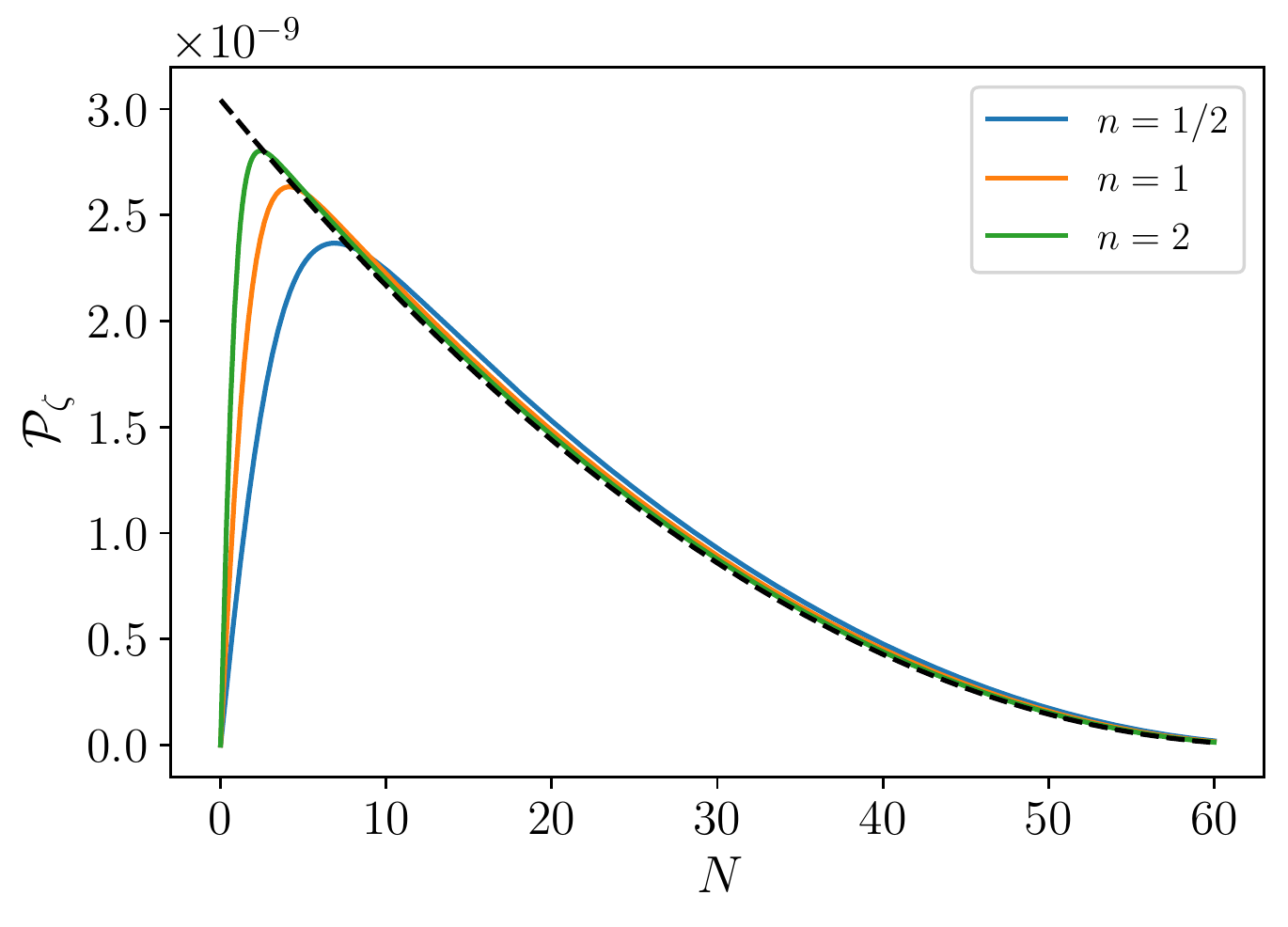}
\caption{Plots of $\big\langle \delta\phi^{(1)\;2} \big\rangle$ (left panel) and $\mathcal{P}_{\zeta}$ (right panel) for the chaotic potential for $n=1/2,1$ and $2$ with the dashed black lines representing the white noise case.}
\label{fig:chaotic_P_zeta_1}
\end{figure}

After computing the power spectrum for stochastic dynamics sourced by white noise, we now turn our attention to colored noise. We have seen that for the general exponential filtering, the noise correlation function turns out to be the following
\begin{equation}
\Big\langle \xi(N_1)\xi(N_2) \Big\rangle_{(n)}=\frac{n}{2}\frac{1}{\cosh\left[ n(N_1-N_2) \right]+1}
\end{equation}
The factor $H^2/4\pi^2$ can been removed in order to preserve the structure of Eq. \eqref{eq:langevin}. The objective, again, would be to compute $\mathcal{P}_{\zeta}$. The situation is now a bit more involved since $\big\langle \xi\delta\phi^{(1)} \big\rangle$ does not reduce to a simple expression. Moreover, it is more expedient to use the $e$-fold variable and the fact that the background evolution for the quadratic potential has an exact expression in slow-roll -- $\phi(N)=\sqrt{\phi_{\text{in}}^2 - 4M_{\text{pl}}^2 N}$. In fact, Eq. \eqref{eq:expec} can be solved analytically in this case. Since,
\begin{equation}
\exp\left[ 2M_{\text{pl}}^{2}\int_{0}^{N}\dd \widetilde{N}\left( \frac{H'}{H} \right)' \right]=\left( \frac{\phi_{\text{in}}^2 - 4M_{\text{pl}}^{2}N}{\phi_{\text{in}}^2} \right)^{1/2}
\end{equation}
we find that
\begin{align}\label{eq:expec2}
\Big\langle \xi\delta\phi^{(1)} \Big\rangle_{(n)}&=\frac{m}{\sqrt{6\pi^2}M_{\text{pl}}}\frac{\phi_{\text{in}}}{2}\left(\frac{\phi_{\text{in}}^2 - 4M_{\text{pl}}^{2}N}{\phi_{\text{in}}^2}  \right)^{-1/2}\int_{0}^{N}\dd N_1\:\frac{n}{2}\frac{1}{\cosh\left[ n(N-N_1) \right]+1}\left( 1-\frac{4M_{\text{pl}}^2}{\phi_{\text{in}}^2}N_1 \right)\nonumber\\
&=\frac{H(\phi_{\text{in}})}{4\pi}\left(1-4\mu^2 N  \right)^{-1/2}\bigg[ \underbrace{\tanh \left(\frac{n}{2}N\right)}_{\substack{\text{small $e$-fold}\\ \text{behavior}}}- \underbrace{\frac{8\mu^2}{n}\ln\left[\cosh \left( \frac{n}{2}N \right)\right]}_{\text{large $e$-fold behavior}}\bigg]
\end{align}
where $\mu=M_{\text{pl}}/\phi_{\text{in}}$ and we clearly see the small and large $e$-fold behavior of the function (early and late times in inflation, respectively). In the asymptotically early inflation time limit ($N\sim 0$)
\begin{equation}
\Big\langle \xi\delta\phi^{(1)} \Big\rangle_{(n)}\simeq\frac{H(\phi_{\text{in}})}{4\pi}\tanh\left( \frac{n}{2}N \right)
\end{equation}
which is, indeed, the correct small $e$-fold behavior. Comparing Eq. \eqref{eq:expec2} with the white noise case $H/4\pi$, we immediately see a point of departure. The terms in the square parentheses all vanish at $N=0$. This, however, is not the case for the white noise case since $H\sim\phi_{\text{cl}}$, being nonzero at $N=0$. The small $e$-fold behavior is dominated by the $\tanh$ function which later becomes congruent with the white noise result. Hence, there is expected to be a deviation for a small range of $e$-folds at the beginning of inflation (as can be seen in Fig. (\ref{fig:chaotic_P_zeta_1})). Further, calculating $\big\langle \delta\phi^{(1)\;2} \big\rangle$, we get
\begin{align}
\Big\langle \delta\phi^{(1)\:2} \Big\rangle_{(n)}&=\frac{H(\phi_{\text{in}})}{4\pi^2}\left(\frac{\phi_{\text{in}}^2 - 4M_{\text{pl}}^{2}N}{\phi_{\text{in}}^2}  \right)^{-1}\int_{0}^{N}\dd N_1 \frac{m}{\sqrt{6}M_{\text{pl}}\phi_{\text{in}}}\left( \phi_{\text{in}}^2-4M_{\text{pl}}^2 N_1 \right)\nonumber\\
&\qquad\qquad\qquad\qquad\qquad\qquad\times\left[ \tanh\left( \frac{n}{2}N_1 \right)-\frac{8M_{\text{pl}}^2}{n\phi_{\text{in}}^2}\ln\left[ \cosh\left( \frac{n}{2}N_1 \right) \right] \right]\nonumber\\
&=\frac{H^2(\phi_{\text{in}})}{4\pi^2}\left(1-4\mu^2 N \right)^{-1}\Bigg\{ \frac{2}{n}\ln\left[ \cosh\left( \frac{n}{2}N \right)\right]\left[ 4\mu^2 N\left( 2\mu^2 N-1 \right) +1\right]\nonumber\\
&-8\mu^4 N^2\left[ \frac{N}{3}+\frac{2}{n}\ln\left( 1+e^{-nN} \right) \right]+\frac{32\mu^4}{n^2}\left[ N\text{Li}_2\left(-e^{-nN}\right)+\frac{1}{n}\text{Li}_3\left(-e^{-nN}\right) \right] \nonumber\\
&\qquad\qquad\qquad\qquad\qquad+\frac{24\mu^4}{n^3}\zeta(3) \Bigg\}
\end{align}
where $\zeta(n)$ is the Riemann zeta function and $\text{Li}_{n}(z)$ are polylogarithm functions, defined by
\begin{equation}
\text{Li}_n(z)=\sum_{k=1}^{\infty}\frac{z^k}{k^n}
\end{equation}
Comparing this with the white noise expression
\begin{equation}
\Big\langle \delta\phi^{(1)\:2} \Big\rangle=\frac{H^2(\phi_{\text{in}})}{4\pi^2}\left(1-4\mu^2 N \right)^{-1}\left(  1-4\mu^2 N+\frac{16}{3}\mu^4 N^2\right)N 
\end{equation}
we, again, observe that the deviation from white noise is only present in the early $e$-fold regime. Finally, computing the power spectrum, we arrive at 
\begin{align}
\mathcal{P}^{(n)}_{\zeta}&=\frac{H^2(\phi_{\text{in}})}{8\pi^2M^2_{\text{pl}}\epsilon_1(\phi_{\text{in}})}\Bigg\{ -8\mu^4\left( 1-\frac{2}{3}\frac{e^{-nN}}{1+e^{-nN}} \right)N^2+\frac{320\mu^4}{9n}N\ln\left( 1+e^{-nN} \right)\nonumber\\
&+\frac{8\mu^2}{n}\left( 4\mu^2 N-1 \right)\ln\left[ \cosh\left( \frac{nN}{2} \right) \right]+\left[ 4\mu^2 N\left( 2\mu^2 N-1 \right) +1\right]\tanh\left( \frac{nN}{2} \right) \Bigg\}
\end{align}
Asymptotically,
\begin{align}
 \mathcal{P}^{(n)}_{\zeta}\simeq\frac{H^2(\phi_{\text{in}})}{8\pi^2M^2_{\text{pl}}\epsilon_1(\phi_{\text{in}})}
\begin{dcases}
\frac{n}{2}N-3n\mu^2 N^2+\left( 4n\mu^4-\frac{n^3}{24} \right)N^3 &\left( 0\lesssim N\lesssim\Delta n \right)\\
16\mu^4 N^2&\left( \text{large }N \right)
\end{dcases}
\end{align}
 
\begin{figure}
\centering
\includegraphics[scale=0.5]{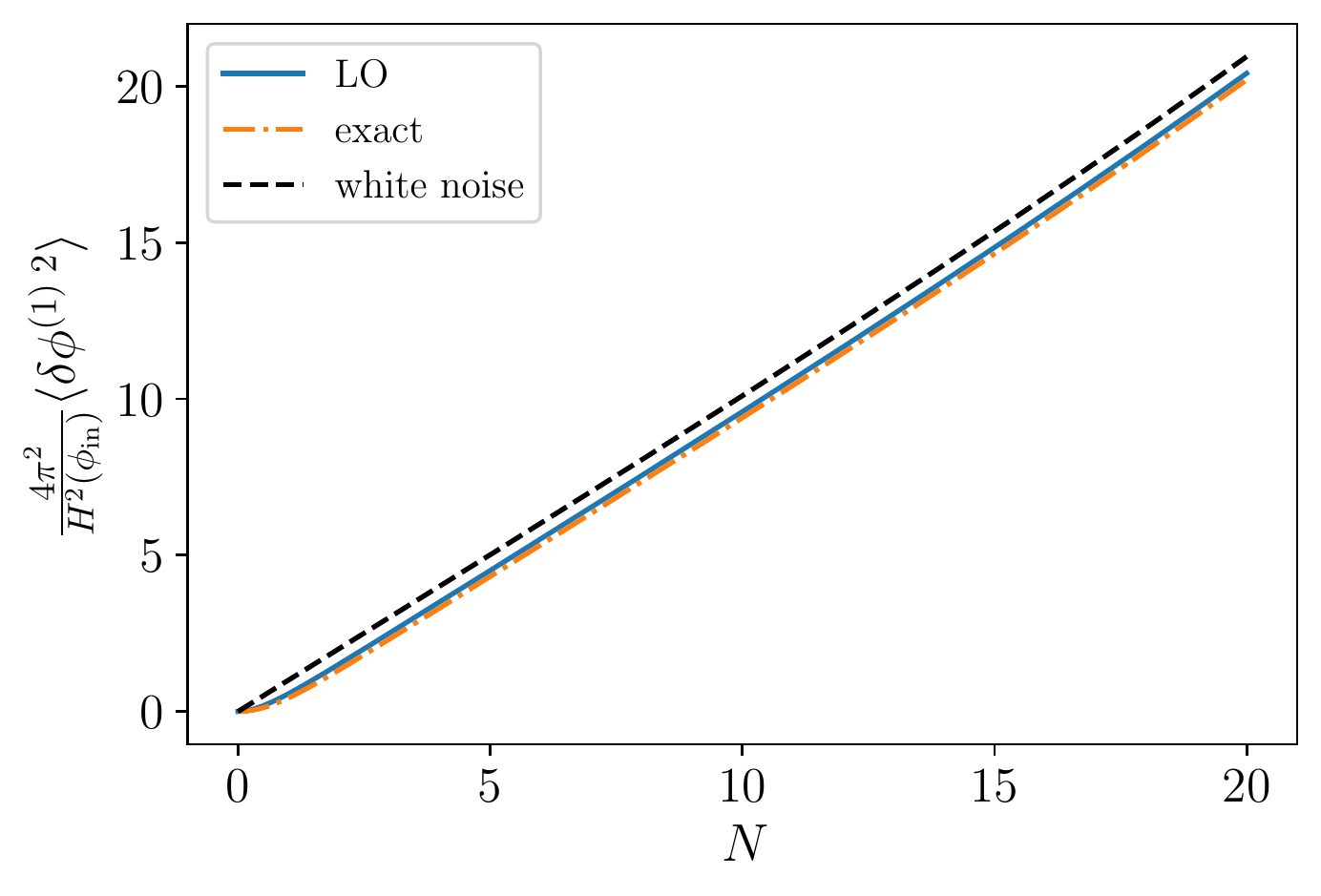}
\includegraphics[scale=0.5]{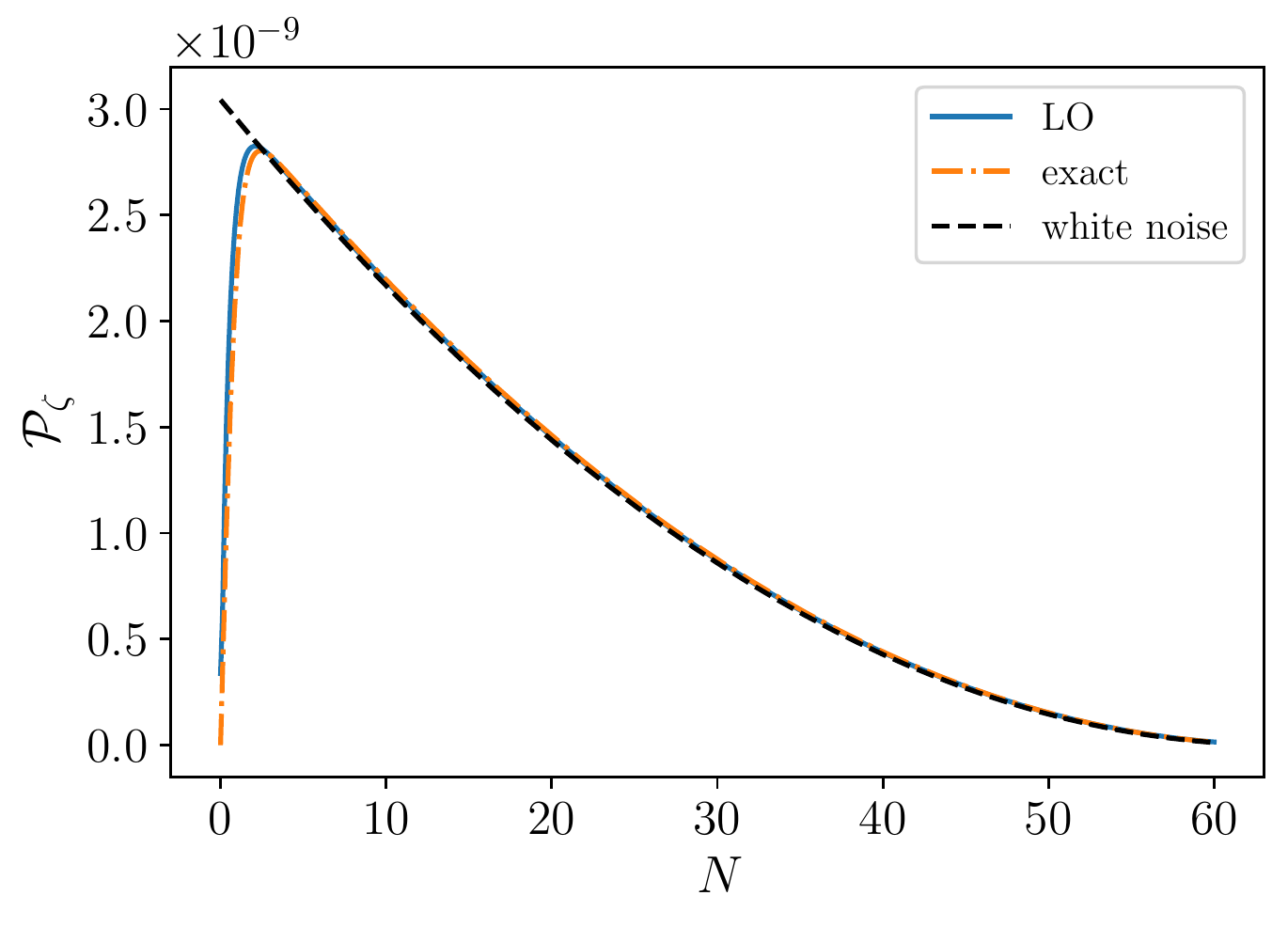}
\caption{Plots of $\big\langle \delta\phi^{(1)\;2} \big\rangle$ (left panel) and $\mathcal{P}_{\zeta}$ (right panel) for the chaotic potential for $n=2$ only. The leading order term, produced by the OU noise, is compared with the exact expressions for $n=2$ (dot dashed) and white noise cases (dashed). The variations between leading order and exact correlation functions are minimal, showing that the OU process is a suitable substitution for numerical simulations.}
\label{fig:chaotic_P_zeta_2}
\end{figure}

The small $e$-fold asymptotics agree with the heuristic derivation in Eq. \eqref{eq:Pzeta_asymp1}. On the other hand, however, the $n$ dependence drops out from the late $e$-fold asymptotics, bearing similarities to the white noise case. This shows that a smooth coarse-graining mostly modifies the small $e$-fold dynamics. These are plotted in Fig. (\ref{fig:chaotic_P_zeta_1}), where we clearly see that the colored noise produces a suppression at the early stages of inflation. This can be understood in terms of $\big\langle \delta\phi^{(1)\;2} \big\rangle$. With $\xi$ correlated for a few $e$-folds at the beginning of inflation, the average variation in the field fluctuation is small since the noise retains some `memory' of its previous states. As a result of this, the ensemble averages of $\delta\phi^{(1)}$ for colored noise are smaller compared to their white noise counterparts. This is all the more evident with smaller values of $n$ -- the smaller the value of $n$, the longer the typical decay time for the correlation function and the longer the suppression lasts. The observable period of inflation is relatively unaffected if the cosmological pivot scale is chosen in such a way that $\mathcal{P}_{\zeta}$ reaches its slow-roll value before its onset. The large $e$-fold $N^2$ behavior is not significant for the purposes of studying inflation since this limit starts taking effect when $N$ is quite large and not the usual $50-60$ $e$-folds that are required. It would also be interesting to study the behavior of $\big\langle \xi\delta\phi^{(1)} \big\rangle$ in an order-by-order expansion since we can recover the OU correlation function at leading order. We study the $n=2$ case (Gaussian filtering) and expand the noise correlation function as a power series in the following way
\begin{align}
\Big\langle \xi(N_1)\xi(N_2) \Big\rangle_{(2)}&=\frac{1}{2\cosh^2(N_1-N_2)}\nonumber\\
&=\underbrace{2e^{-2\Delta N}}_{\text{OU correlation}}\cdot\left( 1-2e^{-2\Delta N}+3e^{-4\Delta N}-\cdot\cdot\cdot \right)\\
&=2e^{-2\Delta N}\sum_{\ell=0}^{\infty}\underbrace{(-1)^\ell (\ell+1)}_{\mathcal{C}_\ell}\left(e^{-2\Delta N}\right)^\ell \label{eq:expansion}
\end{align}
where $\Delta N=N_1-N_2$ and $\mathcal{C}_{n}$ are the coefficients in the expansion. Since this is an expansion about the tail of the two-point correlation function, there will be certain inaccuracies in the small $e$-fold regime. Using this, we can examine the leading order (LO) term in the expansion. Using the LO term in the expansion in Eq. \eqref{eq:expansion}, the integral in Eq. \eqref{eq:expec2} can be expressed as

\begin{align}
\Big\langle \xi\delta\phi^{(1)} \Big\rangle_{(2)}=\frac{H(\phi_{\text{in}})}{4\pi}\left(\frac{\phi_{\text{in}}^2 - 4M_{\text{pl}}^{2}N}{\phi_{\text{in}}^2}  \right)^{-1/2}\int_{0}^{N}\dd N_1\Bigg[&2e^{-2(N-N_1)}\sum_{\ell=0}^{\infty}\mathcal{C}_{\ell}e^{-2\ell(N-N_1)} \nonumber\\
&\times\left( 1-\frac{4M_{\text{pl}}^2}{\phi_{\text{in}}^2}N_1 \right)\Bigg]
\end{align}
These can now be integrated order-by-order. The LO term of $\big\langle \delta\phi^{(1)\;2} \big\rangle$ and $\mathcal{P}_\zeta$ are presented. From Fig. (\ref{fig:chaotic_P_zeta_2}) we can see that leading order, Ornstein-Uhlenbeck noise reproduces the exact results very accurately, accounting for only minor deviations at the small $e$-fold regime (as expected). This shows that a noise correlation function of the form $e^{-2(N-N')}$ can serve as a suitable substitute for the actual noise $\text{sech}^{2}(N-N')$ in the $n=2$ case, the latter being much more difficult to simulate numerically. The same can be shown to be true for other values of $n$ such that 
\begin{equation}\label{eq:correlation_leading_order_general}
\boxed{\Big\langle \xi(N)\xi(N') \Big\rangle_{(n)}\xLongrightarrow{\text{LO}}ne^{-n(N-N')}}
\end{equation}

\subsection{Starobinsky inflation}
Although the chaotic models are some of the simplest realizations of inflation, they predict a scalar-to-tensor ratio, $r$, which is larger than the BICEP2/Kek results of $r<0.06$ when combined with Planck data \cite{Ade:2018gkx,Akrami:2018odb}. As a result, these models are as good as ruled out (although they are exemplary toy models). In this section, we apply the analytical methods to the Starobinsky model which is derived by adding an $R^2$ contribution to the Einstein-Hilbert action \cite{Starobinsky:1979ty,Starobinsky:1980te,Vilenkin:1985md}. The predictions made by this model, the spectral tilt ($n_{s}$) and $r$, are in excellent agreement with observational data for $e$-folds in the range $50<N<60$. The Starobinsky potential takes the following form
\begin{equation}\label{eq:Starobinsky}
V(\phi)=\Lambda^{4}\left( 1-e^{-\alpha\phi/M_{\text{pl}}} \right)^2
\end{equation}
where $\alpha=\sqrt{2/3}$. Thus, the Starobinsky model is a zero-parameter model (much like the chaotic potentials) where $\Lambda$ is solely determined by the CMB normalization. It has been shown that the Higgs inflation reduces to the form in Eq. \eqref{eq:Starobinsky} when transformed to the Einstein frame, hence providing a nice correspondence between the two models \cite{Bezrukov:2007ep,Bezrukov:2008ej,GarciaBellido:2011de}. The Starobinsky model has also been widely studied from the context of supergravity-- for example, no-scale supergravity \cite{Ellis:2013xoa,Ellis:2013nxa,Ellis:2018zya}.\\
\indent Like before, we solve the background evolution in slow-roll. Solving for $\phi$, we arrive at
\begin{equation}
N-N_{\text{in}}=\frac{1}{2\alpha}(x-x_{\text{in}})-\frac{3}{4}\left( e^{\alpha x}-e^{\alpha x_{\text{in}}} \right)
\end{equation}
where $x=\phi/M_{\text{pl}}$. The equation can be inverted by using the Lambert function, $\mathcal{W}(x)$ \cite{Martin:2013tda}. The Lambert function is multivalued and has two real branches. In particular, it can be shown that the inflaton trajectory follows the ``$-1$-branch'' of the Lambert function, $\mathcal{W}_{-1}(x)$. Performing the inversion, the inflaton trajectory can be expressed as
\begin{equation}
x(N)=\frac{1}{\alpha}\left\{ 2\alpha^{2}N+\alpha x_{\text{in}} -e^{\alpha x_{\text{in}}} -\mathcal{W}_{-1}\left[ -\exp\left( 2\alpha^{2}N+\alpha x_{\text{in}}-e^{\alpha x_{\text{in}}} \right) \right] \right\}
\end{equation}
\begin{figure}
\centering
\includegraphics[scale=0.5]{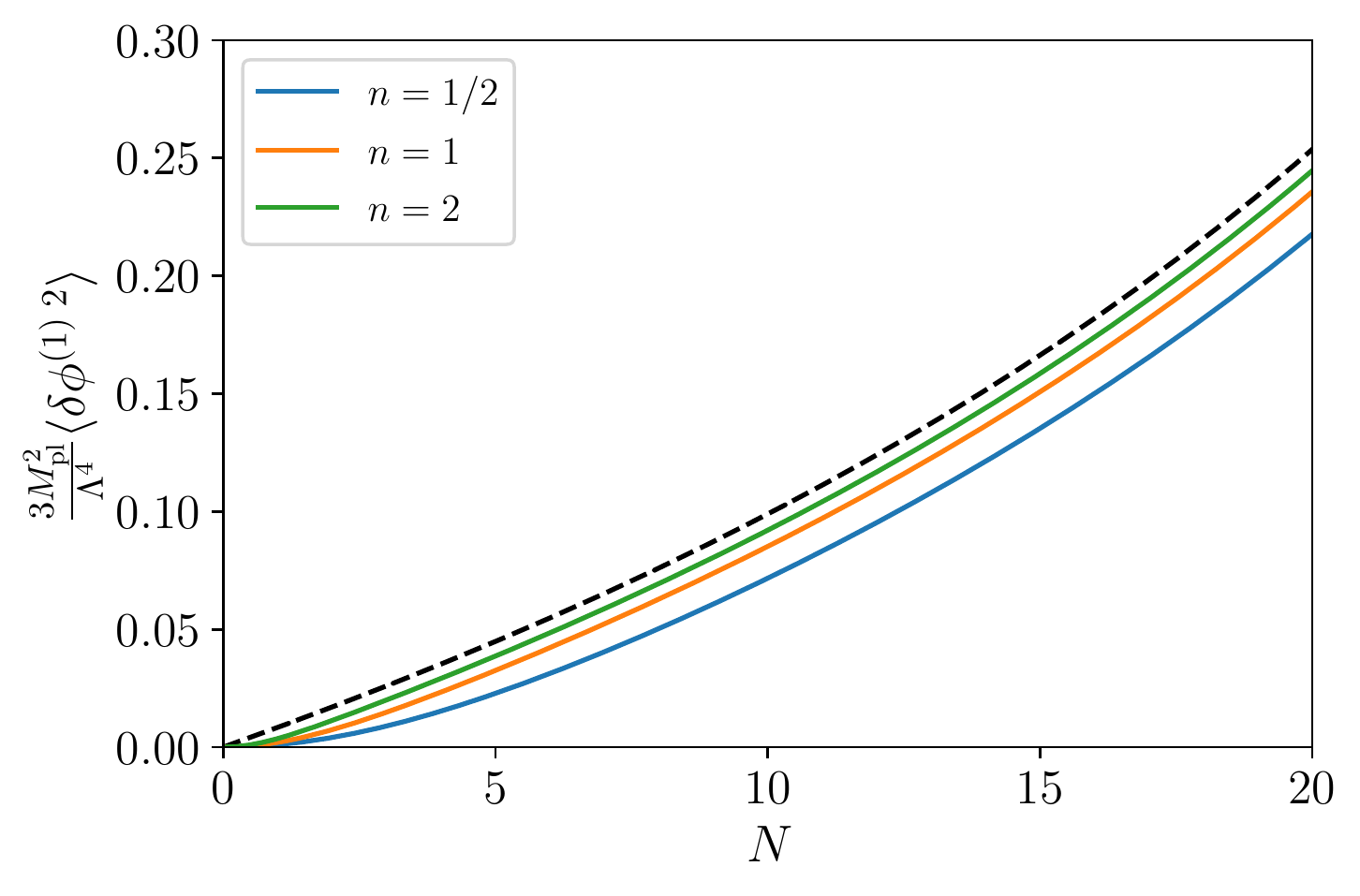}
\includegraphics[scale=0.5]{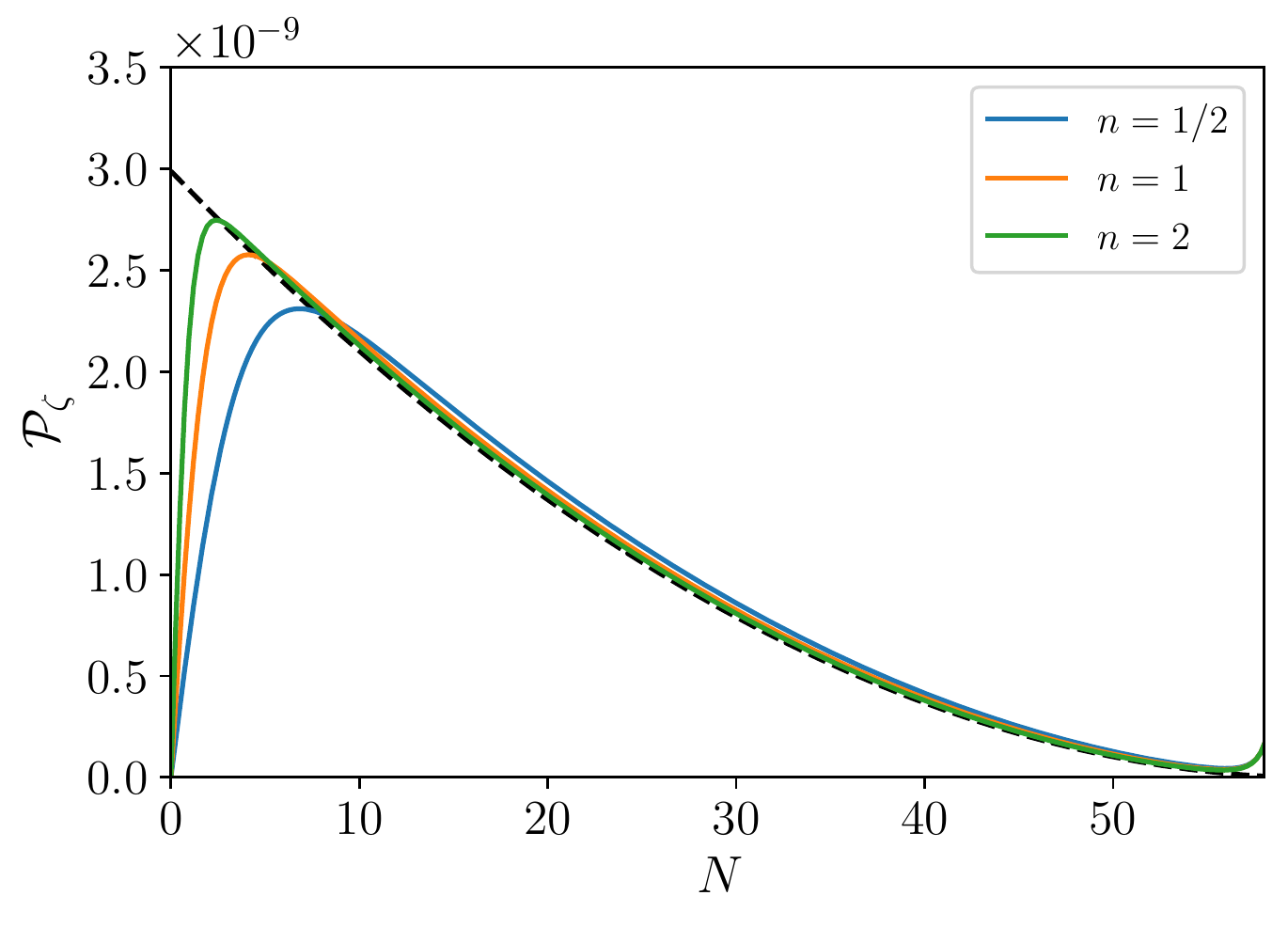}
\caption{Plots of $\big\langle \delta\phi^{(1)\;2} \big\rangle$ (left panel) and $\mathcal{P}_{\zeta}$ (right panel) for the Starobinsky potential for $n=1/2,1$ and $2$ with the dashed black lines representing the white noise case.}
\label{fig:Starobinsky_Pzeta}
\end{figure}
with $N_{\text{in}}=0$. It is easy to see why the inflaton trajectory excursions are along $\mathcal{W}_{-1}$. The domain of $\mathcal{W}_{-1}$ is $[-e^{-1},0)$ and we see that $ -\exp\left( 2\alpha^{2}N+\alpha x_{\text{in}}-e^{\alpha x_{\text{in}}} \right)$ is very small, becoming progressively more negative with increasing $N$.\\
\indent For the Starobinsky potential, we consider $x_{\text{in}}=5.45$ (producing around 60 $e$-folds of inflation) and set $\Lambda=3.4\times 10^{-3}M_{\text{pl}}$. Equation \eqref{eq:expec} can be simplified into the following
\begin{equation}\label{eq:expec3}
\Big\langle \xi\delta\phi^{(1)} \Big\rangle_{(n)}=\frac{e^{-\alpha x(N)}}{1-e^{-\alpha x(N)}}\frac{n}{4\pi}\left( \frac{\Lambda^4}{3M_{\text{pl}}^2} \right)^{1/2}\int_{0}^{N}\dd N'\;\frac{e^{\alpha x(N')}}{\cosh\left[ n(N-N') \right]+1}\left( 1-e^{-\alpha x(N')}\right)^{2} 
\end{equation}
Due to the complicated nature of $x(N)$, we will not attempt to perform analytical calculations. Equation \eqref{eq:expec3} can also be expanded using the methods shown in the previous section, but we do not perform the calculations on the leading order contributions in this section expecting the results from the expansion to be equally valid here. Then,
\begin{align}
\Big\langle \delta\phi^{(1)\;2} \Big\rangle_{(n)}=\frac{e^{-2\alpha x(N)}}{\left( 1-e^{-\alpha x(N)} \right)^2}\frac{n}{4\pi^2}\frac{\Lambda^4}{3 M_{\text{pl}}^2}&\int_{0}^{N}\dd N'\;e^{\alpha x(N')}\left( 1-e^{-\alpha x(N')} \right)^{2}\nonumber\\
&\times\int_{0}^{N'}\dd N''\;\frac{e^{\alpha x(N'')}}{\cosh\left[ n(N'-N'') \right]+1}\left( 1-e^{-\alpha x(N'')} \right)^{2}
\end{align}
In Fig. (\ref{fig:Starobinsky_Pzeta}) we plot $\big\langle \delta\phi^{(1)\;2} \big\rangle$ and $\mathcal{P}_\zeta$ as was done for the chaotic potential. We again observe the characteristic suppression of the power spectrum at small $e$-folds and the time it takes to join with the slow-roll result is dependent on the value of $n$. This shows that this is a feature applicable to any potential -- although the analytical calculations are heavily dependent on the slow-roll approximation.\\
\indent We conclude this section with the recapitulation: the expression of $\big\langle \delta\phi^{(1)\;2} \big\rangle$ in Eq. \eqref{eq:<delphi2>} can be used to first compute the variance of the field fluctuations and, subsequently, the power spectrum for noise two-point correlation functions with finite correlation time. However, the extent to which these can be solved analytically is highly dependent on the complexity of the correlation functions and (or) potentials. For the general class of exponential coarse-graining, the power spectrum can be solved exactly for polynomial potentials of the form $V(\phi)\propto \phi^p$ and the same is likely possible for potentials of the hilltop type. However, more realistic potentials like the Starobinsky are likely to be too complicated for an exact solution. Regardless of whether closed-form expressions are obtainable or not, numerical integration of Eq. \eqref{eq:expec} and \eqref{eq:<delphi2>} will lead to $\mathcal{P}_\zeta$ along with the characteristic, $n$-dependent small $e$-fold suppression. Furthermore, as we have demonstrated, the LO approximation of the correlation function provides a very good fit to the exact results and can be used in simulating SDEs numerically.

\section{$\mathcal{N}$-statistics and the first passage time analysis}\label{sec:N_stats}
We wish to study the statistics of the stochastic number of $e$-folds $\mathcal{N}$ realized during stochastic inflation and derive its moments $(\big\langle \mathcal{N} \big\rangle,\big\langle \mathcal{N}^2 \big\rangle,\cdot\cdot\cdot)$. For such a purpose, we apply the ``first passage time'' analysis which will allow us to compute the aforementioned moments of the number of $e$-folds. This was applied to the white noise case in Refs. \cite{Vennin:2015hra,Assadullahi:2016gkk,Vennin:2016wnk} where the $\delta\mathcal{N}$ formalism was used to compute $\mathcal{P}_{\zeta}$ for both single and multifield stochastic inflationary scenarios.\\
 \indent Formulating the first passage problem requires a Fokker-Planck equation (FPE) describing the stochastic process. However, we know that FPEs are applicable for Markov processes, of which the Wiener process (white noise) is an example. Colored noise processes, in general, are non-Markovian since the presence of a finite correlation time ruins the Markov property. Nevertheless, approximate FPEs can be derived for stochastic processes in which the correlation time is much smaller than the time-scale of the system \cite{PhysRevA.33.467,PhysRevA.39.6094,doi:10.1063/1.1860471}. In the following sections, the first passage problem will be formulated for the leading order noise correlation function derived earlier for the case $n=2$. For simplicity, we focus on the chaotic potential. In what follows, we study the average $e$-folds realized during inflation $\big\langle \mathcal{N} \big\rangle$ for such a quadratic potential in two ways
 \begin{enumerate}
    \item The first passage problem is formulated which directly leads to a differential equation which relates $\big\langle \mathcal{N}^n \big\rangle$ to $\big\langle \mathcal{N}^{n-1} \big\rangle$. The ensuing differential equation can be numerically integrated after imposing suitable boundary conditions and $\big\langle \mathcal{N} \big\rangle$ can be calculated accordingly.
    \item The first passage problem is transformed into an equivalent one by means of the characteristic function. It will be shown that a solution of the characteristic function can be obtained in the form of Gauss hypergeometric functions and, after certain expansions, an approximate expression for $\big\langle \mathcal{N} \big\rangle$ can be obtained.
    \item The results from the two aforementioned methods are finally compared.
 \end{enumerate}
\subsection{Fokker-Planck equation and approximates for colored noise}
The Fokker-Planck equation describes the evolution of the conditional probability density of some random process. In the case of stochastic inflation, this is the probability that the inflaton is at some field position $\phi(N)$ given that is started out from $\phi_{\text{in}}(N_{\text{in}})$. The inflaton-Langevin equation is expressed as
\begin{equation}
\frac{\dd\phi}{\dd N}=-2M^2_{\text{pl}}\frac{H'}{H}+\frac{H}{2\pi}\xi
\end{equation}
When there is white noise, $\big\langle \xi(N_1)\xi(N_2)\big\rangle=\delta(N_1-N_2)$, the evolution of the probability distribution is encoded through the FPE as follows
\begin{align}
\frac{\partial}{\partial N}P(\phi,N|\phi_{\text{in}},N_{\text{in}})&=\frac{\partial}{\partial \phi}\left[ 2M^2_{\text{pl}}\frac{H'}{H} P(\phi,N|\phi_{\text{in}},N_{\text{in}})\right]+\frac{\partial^2}{\partial\phi^2}\left[ \frac{H^2}{8\pi^2}P(\phi,N|\phi_{\text{in}},N_{\text{in}}) \right]\\
&=-\frac{\partial}{\partial \phi}\left[\mu(\phi)P(\phi,N|\phi_{\text{in}},N_{\text{in}})\right]+\frac{1}{2}\frac{\partial^2}{\partial\phi^2}\left[ \sigma^2(\phi)P(\phi,N|\phi_{\text{in}},N_{\text{in}}) \right]\label{eq:FPE_0}
\end{align}
or, in operator notation
\begin{equation}
\frac{\partial}{\partial N}P(\phi,N|\phi_{\text{in}},N_{\text{in}})=\mathcal{L}_{\text{FP}}(\phi)\cdot P(\phi,N|\phi_{\text{in}},N_{\text{in}})
\end{equation}
In Eq. \eqref{eq:FPE_0}, $\mu(\phi)$ and $\sigma(\phi)$ are the drift and diffusion terms in standard notation, commonly referring to the deterministic and random parts of the SDE. When the noise becomes correlated (colored), it turns out that an FPE cannot be derived since it loses the Markov property. However, approximate FPEs can be derived for noise where the correlation function decays very rapidly. For stochastic inflation derived from the general class of exponential coarse-graining, it is seen that, at leading order, the noise correlation function has the following form
\begin{equation}
\Big\langle \xi(N_1)\xi(N_2) \Big\rangle\simeq 2e^{-2|N_1-N_2|}
\end{equation}
For the time-scale of the system being 50-60 $e$-folds for inflation, this correlation function decays very rapidly. The approximate FPE can be derived using the functional calculus approach where the noise modeled as a Gaussian random field given by the following probability density functional \cite{PhysRevA.33.467}
\begin{equation}
P[\xi]=N\exp\left[ -\frac{1}{2}\int\dd N \int\dd N'\xi(N)\xi(N')\mathcal{K}(N-N') \right]
\end{equation}
with the normalization given by 
\begin{equation}
N^{-1}=\int\mathcal{D}\xi\exp\left[ -\frac{1}{2}\int\dd N \int\dd N'\xi(N)\xi(N')\mathcal{K}(N-N') \right]
\end{equation}
where $\mathcal{K}(N-N')$ is the inverse of the colored noise correlation function. In Refs. \cite{PhysRevA.33.467,PhysRevA.39.6094}, the authors used a noise correlation function of the form $\big\langle \xi(t)\xi(s) \big\rangle\simeq \tau^{-1}e^{-\frac{|t-s|}{\tau}}$, where $\tau$ represents the small correlation time. Since $e$-folds is a dimensionless quantity, we use the $\tau=1/2$ (which is the prefactor in the exponential function). Hence, using a small-$\tau$ expansion, the approximate FPE can be expressed as follows
\begin{equation}
\frac{\partial}{\partial N}P(\phi,N|\phi_{\text{in}},N_{\text{in}})=-\frac{\partial}{\partial \phi}\left[\mu(\phi)P(\phi,N|\phi_{\text{in}},N_{\text{in}})\right]+\frac{1}{2}\frac{\partial^2}{\partial\phi^2}\left[ \widetilde{\sigma}^2(\phi)P(\phi,N|\phi_{\text{in}},N_{\text{in}}) \right]\label{eq:FPE}
\end{equation}
where the new diffusion term is given by
\begin{equation}
\widetilde{\sigma}^2(\phi)=\sigma(\phi)\left[ \sigma(\phi)+\frac{1}{2}\sigma(\phi)\frac{\partial\mu(\phi)}{\partial\phi}-\frac{1}{2}\frac{\partial\sigma(\phi)}{\partial\phi}\mu(\phi) \right]=\sigma^2(\phi)+\text{corrections}
\end{equation}
We see that new diffusion term receives corrections from derivative contributions arising from $\sigma'(\phi)$ and $\mu'(\phi)$. If these derivatives are small and subdominant, the diffusion term is approximately similar to the white noise case. 

\subsection{First passage time and $e$-fold moments}
In order to compute the statistics of the number of $e$-folds realized during stochastic inflation, one must specify the boundaries in field space. For a simple potential like $\phi^2$, the trajectory of the inflaton is one where it simply rolls down the potential from a super Planckian field value to the point where inflation is terminated. Since inflation ends when $\epsilon_1 =1$, the endpoint for the chaotic potential is $\phi_{\text{end}}=\sqrt{2}M_{\text{pl}}$. On the other hand, the large-field endpoint is kept arbitrary for now, only being labelled as $\phi_\text{uv}$. The number of $e$-folds realized between $\phi_\text{uv}$ and $\phi_\text{end}$ is labelled as $\mathcal{N}$, which is a quantity dependent on the particular realization of the stochastic process.\\
\indent In Appendix \ref{sec:first_passage_deriv}, the first passage time formalism is derived, with which it can be shown that the moments of the $e$-fold satisfy a hierarchical relationship given by the following\footnote{The $\phi_\text{in}$ for the FPE has been relabeled to $\phi$. However, the interpretation has not been changed since the FPE still contains derivatives with respect to the initial $e$-fold and field values.}

\begin{equation}\label{eq:first_passage}
\underbrace{\left( \mu(\phi)\frac{\partial}{\partial\phi}+\frac{1}{2}\widetilde{\sigma}^2(\phi)\frac{\partial^2}{\partial\phi^2} \right)}_{\mathcal{L}^{\dagger}_{\text{FP}}(\phi)}\Big\langle \mathcal{N}^n \Big\rangle(\phi)=-n\Big\langle \mathcal{N}^{n-1} \Big\rangle(\phi)
\end{equation}
where $\mathcal{L}^{\dagger}_{\text{FP}}(\phi)$ is the operator satisfying the adjoint FPE.\footnote{This is also known as the backward Kolmogorov equation. Backward integration is more convenient when computing first crossing probabilities, hence its use in first passage problems.} The fact that $\mathcal{L}^{\dagger}_{\text{FP}}(\phi)$ is the adjoint of $\mathcal{L}_{\text{FP}}(\phi)$ can be shown by $\big\langle f,\mathcal{L}_\text{FP}(g) \big\rangle=\int\dd\phi f(\phi)\mathcal{L}_\text{FP}(g(\phi))=\int\dd\phi \mathcal{L}^\dagger_\text{FP}(f(\phi))g(\phi)=\big\langle \mathcal{L}^\dagger_\text{FP}(f),g \big\rangle$. Equation \eqref{eq:first_passage} can be solved iteratively for $n\geq 1$ using $\big\langle \mathcal{N}^0 \big\rangle=1$. Hence, the mean number of $e$-folds is given by the following differential equation
\begin{equation}\label{eq:<N>}
\mathcal{L}^\dagger_\text{FP}(\phi)\cdot\Big\langle\mathcal{N}\Big\rangle=-1
\end{equation}
\begin{figure}
    \centering
    \begin{tikzpicture}[scale=0.75, transform shape]
        \fill[red, opacity=0.25] (6.5,0) rectangle (7.5,6);
        \fill[gray, opacity=0.25] (1.36,0) rectangle (-1,6);
        \node at (-2,3){\begin{tabular}{c}Absorbing boundary\\$P(\phi_\text{end},\mathcal{N}|\phi_\text{in},N)=0$\end{tabular}};
        \node at (8.25,3){\begin{tabular}{c}Reflecting boundary\\$\frac{\partial}{\partial\phi_\text{in}}P(\phi,N|\phi_\text{in},N)\bigg\lvert_{\phi=\phi_\text{uv}}=0$\end{tabular}};
        \draw[very thick, ->] (-1,0) -- (8,0)node[right]{$\phi/M_{\text{pl}}$};
        \draw[very thick, ->] (0,-1) -- (0,6)node[above]{$V(\phi)$};
        \draw[<->, thick] (1.36,-1) -- (6.5,-1)node[midway,below]{$\mathcal{N}\:e\text{-folds}$};
        \draw[line width=1.5mm, green] (6.5,5) to[out=180,in=70] (2,3);
        \draw[line width=1.5mm, green] (2,3) to[out=250,in=66] (1.36,1);
        \draw[ultra thick, blue] (7.5,5) to[out=180,in=70] (2,3);  
        \draw[ultra thick, blue] (2,3) to[out=250,in=0] (0,0);
        \fill[color=red] (6.5,5) circle (0.1);
        \draw[thick, dashed] (6.5,5) -- (6.5,0) node[below]{$\phi_\text{uv}\gg \phi_\text{in}$};
        \fill[color=red] (1.36,1) circle (0.1);
        \draw[thick, dashed] (1.36,1) -- (1.36,0) node[below]{$\phi_{\text{end}}$};
    \end{tikzpicture}
    \caption{An illustration of the absorbing and reflecting boundary conditions at $\phi_\text{end}$ and $\phi_\text{uv}$. The green highlighted region corresponds to the inflationary excursion where the inflaton can exist inside the potential $V(\phi)$ while the shaded regions are areas of exclusion.}
    \label{fig:BC_schema}
\end{figure}
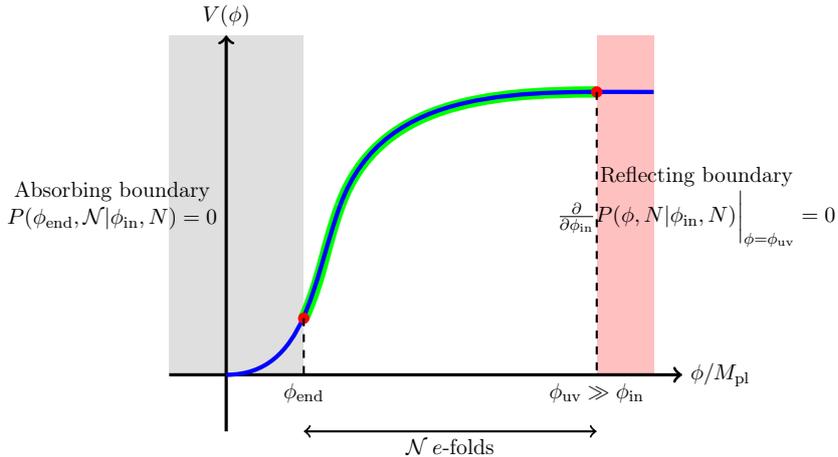
This can be solved via direct integration, with the result expressible for general $\mu(\phi)$ and $\widetilde{\sigma}(\phi)$. However, since Eq. \eqref{eq:first_passage} is a boundary value problem, we need to specify boundary conditions. These can be set as follows (see Fig. (\ref{fig:BC_schema}) for an illustration)
\begin{enumerate}
\item Since inflation terminates at $\phi_\text{end}$, we require that the inflaton be removed from the system. Such a requirement leads to an \textit{absorbing} boundary condition, given by
\begin{equation}
P(\phi_\text{end},\mathcal{N}|\phi_\text{in},N_\text{in})=0
\end{equation}
applicable for all initial conditions. This, in turn, leads to $\big\langle \mathcal{N}^n \big\rangle(\phi_\text{end})=0$.
\item On the other end, one can similarly impose an absorbing boundary at $\phi_\text{uv}$. However, a more well-posed boundary would be one that prevents the inflaton from exploring arbitrarily large field values and, when it reaches a cut-off, bounces back. This is called a \textit{reflecting} boundary and it amounts to stating the following
\begin{equation}
\frac{\partial}{\partial \phi_\text{in}}P(\phi,N|\phi_\text{in},N_\text{in})\bigg\lvert_{\phi=\phi_\text{uv}}=0
\end{equation}
which follows from imposing a vanishing probability current at the boundary. This translates to $\partial_{\phi_\text{in}}\big\langle \mathcal{N}^n \big\rangle(\phi_\text{uv})=0$.
\end{enumerate}
We proceed with the analytical solution of Eq. \eqref{eq:<N>}. Transforming it to first order with the change of variable $\partial_\phi\big\langle \mathcal{N} \big\rangle=\left\langle \mathcal{M} \right\rangle$, we have
\begin{align}
\Big\langle \mathcal{M} \Big\rangle(\phi)&=\frac{\mathcal{C}_1}{ \Psi(\phi)}-\frac{2}{\Psi(\phi)}\int_{\phi_\text{uv}}^{\phi}\dd\phi_2\frac{\Psi(\phi_2)}{\widetilde{\sigma}^2(\phi_2)}
\end{align}
where 
\begin{equation}
\Psi(\phi)=\exp\left[ \int_{\phi_\text{uv}}^{\phi}\frac{2\mu(\phi_1)}{\widetilde{\sigma}^2(\phi_1)} \dd\phi_1\right]
\end{equation}
The constant of integration $\mathcal{C}_1$ is determined by demanding the reflecting boundary condition at $\phi_\text{uv}$, which is $\partial_\phi\big\langle \mathcal{N} \big\rangle|_{\phi_\text{uv}}=\big\langle \mathcal{M} \big\rangle|_{\phi_\text{uv}}=0$. With the reflecting boundary condition at $\phi_\text{uv}$, $\mathcal{C}_1=0$. Upon a final integration, where the second integration constant can be set to zero by imposing the absorbing boundary condition at $\phi_\text{end}$, the average $e$-fold can be expressed as
\begin{equation}\label{eq:efold_average}
\Big\langle \mathcal{N} \Big\rangle(\phi)=2\int_{\phi_\text{end}}^{\phi}\dd\phi_1\int_{\phi_\text{uv}}^{\phi_1}\dd\phi_2\frac{\Psi(\phi_2)}{\Psi(\phi_1)}\frac{1}{\widetilde{\sigma}^2(\phi_2)}
\end{equation}
Equation \eqref{eq:efold_average} gives us the ensemble average of the number of $e$-folds realized in the interval $\phi_\text{end}\leq\phi\lesssim\phi_\text{uv}$ given that there are reflecting and absorbing boundary conditions at $\phi_\text{uv}$ and $\phi_\text{end}$. In general, knowledge of the $n$-th solution of Eq. \eqref{eq:first_passage} allows us compute the $(n+1)$-th solution and, hence, construct a hierarchy of moments. In general,
\begin{equation}\label{eq:efold_nth_moments}
\Big\langle \mathcal{N}^n \Big\rangle(\phi)=2n\int_{\phi_\text{end}}^{\phi}\dd\phi_1\int_{\phi_\text{uv}}^{\phi_1}\dd\phi_2\frac{\Psi(\phi_2)}{\Psi(\phi_1)}\frac{\Big\langle \mathcal{N}^{n-1} \Big\rangle(\phi_2)}{\widetilde{\sigma}^2(\phi_2)}
\end{equation}
\subsection{Characteristic function approach for $e$-fold moments}
The expression for the $n$-th moment of $\mathcal{N}$ is solved either as a series of coupled differential equations or through iterated integrals, which can quickly become cumbersome. Alternatively, the moments of $\mathcal{N}$ can be calculated by making use of the characteristic function \cite{Pattison:2017mbe}. Defining the characteristic function as $\chi(k,\phi)=\sum_{n=0}^{\infty}(ik)^n /n! \big\langle \mathcal{N}^n \big\rangle$, it can can substituted into Eq. \eqref{eq:first_passage} to obtain the following differential equation
\begin{equation}\label{eq:FPE_characteristic}
\left( \mu(\phi)\frac{\partial}{\partial\phi}+\frac{1}{2}\widetilde{\sigma}^2(\phi)\frac{\partial^2}{\partial\phi^2} +ik\right)\chi(k,\phi)=0
\end{equation}
The advantage of this approach is that the hierarchy of coupled differential equations has been reduced to an uncoupled one, with the following boundary conditions
\begin{equation}
\chi(k,\phi_\text{end})=1,\;\;\;\;\;\frac{\partial}{\partial \phi}\chi(k,\phi)\bigg\lvert_{\phi_\text{uv}}=0
\end{equation}
With the knowledge of the characteristic function, the moments of the $e$-fold distribution can be calculated at all orders
\begin{equation}
\Big\langle \mathcal{N}^n \Big\rangle=\frac{1}{i^n}\frac{\partial^n}{\partial k^n}\chi(k,\phi)\bigg\lvert_{k=0}
\end{equation}
For the quadratic potential influenced by white noise, it can be shown that Eq. \eqref{eq:FPE_characteristic} admits closed-form solutions in the form of hypergeometric functions $_{1}F_{1}$ \cite{Pattison:2017mbe}. In our case, with the corrections to the diffusion term, it is possible to obtain the following analytical solution. 
\begin{align}
\chi(k,\phi)&=\mathcal{C}^{(1)}\: _{2}F_1\left( -\frac{1}{4}-\frac{\alpha(k)}{4},-\frac{1}{4}+\frac{\alpha(k)}{4},\frac{1}{2}-\frac{\beta}{2};-\frac{\phi^2}{2M^2_\text{pl}} \right)\nonumber\\
&+\mathcal{C}^{(2)}\:\left( \frac{\phi}{2M_\text{pl}} \right)^{1 +\beta}\: _{2}F_1\left( \frac{1}{4}+\frac{\beta}{2}-\frac{\alpha(k)}{4},\frac{1}{4}+\frac{\beta}{2}+\frac{\alpha(k)}{4},\frac{3}{2}+\frac{\beta}{2};-\frac{\phi^2}{2M^2_\text{pl}} \right)\\
&=\mathcal{C}^{(1)}\chi^{(1)}(k,\phi)+\mathcal{C}^{(2)}\chi^{(2)}(k,\phi)
\end{align}
\begin{figure}
\centering
\includegraphics{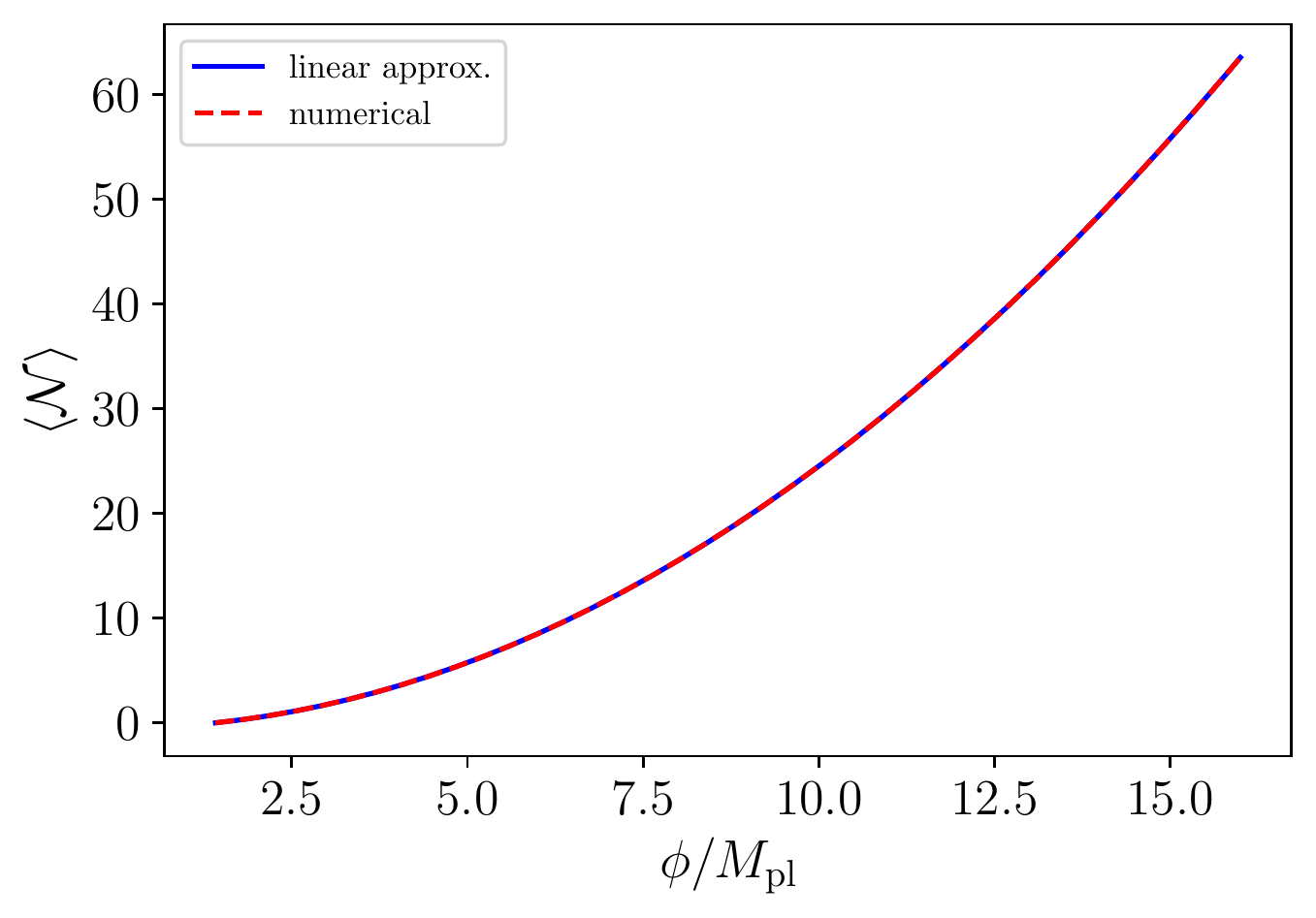}
\caption{Average $e$-folds for the quadratic potential between $\sqrt{2}M_\text{pl}\leq\phi\leq 16M_\text{pl}$ with the dashed red line being produced by numerically solving Eq. \eqref{eq:first_passage} with the reflecting and absorbing boundary conditions.}
\label{fig:ave_efold}
\end{figure}
where $_{2}F_{1}(a,b,c;z)$ are Gauss hypergeometric functions \cite{abramowitz+stegun,822801} and $\alpha(k)=\sqrt{1-4ik\beta}$ and $\beta=\frac{48\pi^2 M^2_\text{pl}}{m^2}$. It can be seen that $\chi^{(2)}(k,\phi)$ produces $\big\langle \mathcal{N} \big\rangle<0$, which is nonphysical. Hence, we set $\mathcal{C}^{(2)}=0$. Then, using the fact that $\chi(k,\phi_\text{end})=1$, the final solution can be expressed as
\begin{equation}
\chi(k,\phi)=\frac{_{2}F_1\left( -\frac{1}{4}-\frac{\alpha(k)}{4},-\frac{1}{4}+\frac{\alpha(k)}{4},\frac{1}{2}-\frac{\beta}{2};-\frac{\phi^2}{2M^2_\text{pl}} \right)}{_{2}F_1\left( -\frac{1}{4}-\frac{\alpha(k)}{4},-\frac{1}{4}+\frac{\alpha(k)}{4},\frac{1}{2}-\frac{\beta}{2};-\frac{\phi^2_\text{end}}{2M^2_\text{pl}} \right)}
\end{equation}
Hypergeometric functions can be notoriously difficult to evaluate when the parameters $a,b$ and $c$ become large. In our present case, only $c$ is large since $\beta\propto m^{-2}$. However, if all parameter are large, the situation can be remedied by performing the Euler transformation on $_{2}F_{1}$, which can be expressed as
\begin{equation}\label{eq:Euler_transform}
_{2}F_{1}(a,b,c;z)=(1-z)^{c-a-b}\:_{2}F_{1}(c-a,c-b,c;z)
\end{equation}
The identity in Eq. \eqref{eq:Euler_transform}, thus, reduces the hypergeometric function from one with three large parameters ($a,b$ and $c$) to only one ($c$). For our solution, since only $c$ is large, the solution can be simplified using the power series expansion of $_{2}F_{1}$, evaluating it up to linear order
\begin{align}
_{2}F_{1}(a,b,c;z)&=\sum_{n=0}^{\infty}\frac{(a)_n (b)_n}{(c)_n}\frac{z^n}{n!}\nonumber\\
                      &\simeq 1 + \frac{ab}{c}z
\end{align}
where the quantity $(a)_n = a(a+1)\cdot\cdot\cdot (a+n-1)$ is the Pochhammer symbol, representing an ascending factorial. This expansion is technically not correct for our particular scenario since the power series expansion is valid for $|z|<1$. However, we recall that $c\sim\beta\sim\mathcal{O}(10^{13})$. Using a small range of $e$-folds $\sqrt{2}M_\text{pl}\leq\phi\leq 16M_\text{pl}$, we expect this expansion to be valid. Hence, in this approximation, the average $e$-fold is given by
\begin{equation}\label{eq:ave_efold_linear}
\Big\langle \mathcal{N} \Big\rangle_\text{colored}(x)\simeq\frac{x^2}{4}\frac{1-\frac{x_\text{end}^2}{x^2}}{1-\frac{m^2}{48\pi^2 M^2_\text{pl}}},\;\;\;\;\;\;\;\left(x=\frac{\phi}{M_\text{pl}}\right)
\end{equation}
The average $e$-folds $\big\langle \mathcal{N} \big\rangle$ for the chaotic potential are plotted in Fig. (\ref{fig:ave_efold}). We observe that, within the range of field values used, the linear approximation using Eq. \eqref{eq:ave_efold_linear} agrees very well with the numerical results obtained from solving Eq. \eqref{eq:<N>}. Moreover, we can compare our results with those obtained for white noise in Ref. \cite{Pattison:2017mbe}. In the notation used in this work, the characteristic function can be written as 
\begin{equation}\label{eq:chi_white}
    \chi_{\text{white}}(k,\phi)=\left( \frac{\phi^2}{\phi^2_{\text{end}}} \right)^{\frac{1-\alpha(k)}{2}}\frac{_{1}F_{1}\left( -\frac{1}{4}+\frac{\alpha(k)}{4},1+\frac{\alpha(k)}{2};-\frac{\beta M^2_\text{pl}}{\phi^2} \right)}{_{1}F_{1}\left( -\frac{1}{4}+\frac{\alpha(k)}{4},1+\frac{\alpha(k)}{2};-\frac{\beta M^2_\text{pl}}{\phi^2_\text{end}} \right)}
\end{equation}
\begin{figure}
    \centering
    \includegraphics{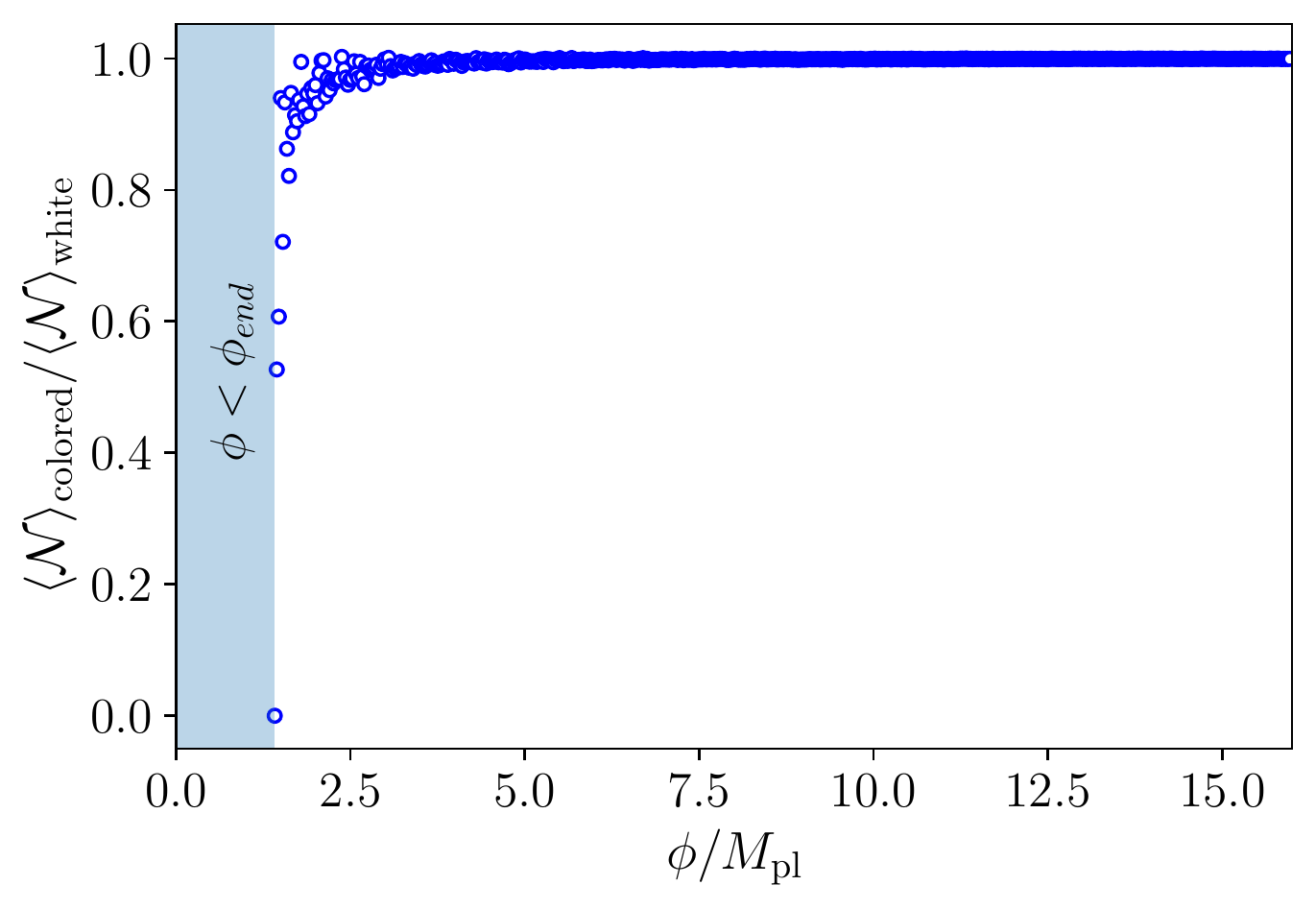}
    \caption{A comparison of the average $e$-folds for the quadratic potential between $\sqrt{2}M_\text{pl}\leq\phi\leq 16M_\text{pl}$ produced during stochastic inflation driven by the exponential correlation and uncorrelated noise. The blue, shaded region corresponds to $\phi<\phi_\text{end}$. The quantity $\langle \mathcal{N} \rangle_\text{white}$ has been computed by direct differentiation of Eq. \eqref{eq:chi_white}.}
    \label{fig:efold_comparison}
\end{figure}
where $_{1}F_{1}$ is the confluent hypergeometric function. The ratio between the $e$-folds $\langle \mathcal{N} \rangle_\text{colored}/\langle \mathcal{N} \rangle_\text{white}$ is plotted in Fig. (\ref{fig:efold_comparison}), where we see that there is negligible difference. This can be understood by observing the fact that the diffusion term in the approximate FP equation $\widetilde{\sigma}(\phi)$ receives very small corrections. In fact, we can show that (by scaling arguments)
\begin{equation}
    \frac{\widetilde{\sigma}^2-\sigma^2}{\sigma^2}\equiv\frac{\Delta\sigma^2}{\sigma^2}\sim \frac{m^2}{m^2\phi^2/M_\text{pl}^2}\sim\frac{M^2_\text{pl}}{\phi^2}
\end{equation}
The only observable deviation occurs when $\phi_\text{in}$ is very small, such that the inflationary excursion consists of very low $e$-folds. We may surmise that this is due to the effects of the finite correlation time at the beginning of inflation. It may happen that, for larger $\phi_\text{in}$, the inflaton trajectory relaxes to that of one described by white noise after an initial disturbance by the colored noise (provided it decays sufficiently quickly). On the other hand, for very small $\phi_\text{in}$, the inflaton trajectory might not have enough time to relax back to that of white noise. This can manifest itself in the form of the very small deviations when $\phi_\text{in}$ is close to $\phi_\text{end}$.  

\section{Numerical implementation of colored noise}\label{sec:colored_noise_numerical}
In this chapter we provide a description of how to simulate a stochastic process governed by colored noise. In particular, we study the $n=2$ case where the noise correlation function is $\big\langle \xi(N)\xi(N') \big\rangle= 2\:\text{sech}^2(N-N')$. Since we have already seen that the leading order, exponential form of the exact correlation function reproduces the coarse-grained dynamics very well, we use it in numerical simulations. For general $n$, the expression in Eq. \eqref{eq:correlation_leading_order_general} can be used for numerical calculations. The procedure for discretizing the SDEs for any general value of $n$ can be found in Appendix \ref{eq:OU_process_derivation}.
\subsection{Stochastic dynamics with Ornstein-Uhlenbeck noise correlation}
A simple example of a stochastic process that contains a finite time correlation is to consider the following
\begin{equation}\label{eq:stochastic_example}
\frac{\dd\xi}{\dd t}=-\frac{1}{\tau}\xi+\frac{1}{\tau}\eta(t)
\end{equation}
where $\eta(t)$ is Gaussian white noise with $\big\langle \eta(t) \big\rangle=0$ and $\big\langle \eta(t_{1})\eta(t_{2}) \big\rangle=N\delta(t_{1}-t_{2})$. For the familiar reader, this is nothing but the motion of a Brownian particle with a relaxation time $\tau$. Equation \eqref{eq:stochastic_example} can be directly integrated to yield the following correlation function of $\xi$
\begin{align}
\Big\langle \xi(t_{1})\xi(t_{2}) \Big\rangle&=\frac{N}{\tau^2}\int_{t_{\text{in}}}^{\text{min}(t_{1},t_{2})}\dd t'e^{(2t'-t_{1}-t_{2})/\tau}\nonumber\\
&=\frac{N}{2\tau}\left[ e^{-|t_{1}-t_{2}|/\tau}-e^{(2t_{\text{in}}-t_{1}-t_{2})/\tau} \right]
\end{align}
where we have assumed that $\xi(t_{\text{in}})=0$ for some initial time. A complete derivation can be found in Appendix \ref{eq:OU_process_derivation}. In the late time limit, the correlation function reduces to $\Big\langle \xi(t_{1})\xi(t_{2}) \Big\rangle=N(2\tau)^{-1}e^{-|t_{1}-t_{2}|/\tau}$ with the system now being described by some finite correlation time, $\tau$. Of course, in the limit $\tau\rightarrow0$, we recover a Dirac $\delta$ correlation. Now, if a system were to be influenced by $\xi(t)$ via some Langevin equation 
\begin{equation}\label{eq:stochastic_example2}
\frac{\dd x}{\dd t}=\mu(x)+\sigma(x)\xi(t)
\end{equation}
a numerical solution would imply simultaneously solving Eq. \eqref{eq:stochastic_example} and \eqref{eq:stochastic_example2}. More succinctly, we solve the following coupled SDEs \cite{Milstein,doi:https://doi.org/10.1002/9780470141489.ch4}
\begin{align}
\dd \xi&=-\frac{1}{\tau}\xi(t)\dd t+\frac{1}{\tau}\dd W(t)\\
\dd x&=\mu(x)\dd t+\sigma(x)\xi(t)\dd t
\end{align}
\begin{figure}[t]
\centering
\includegraphics{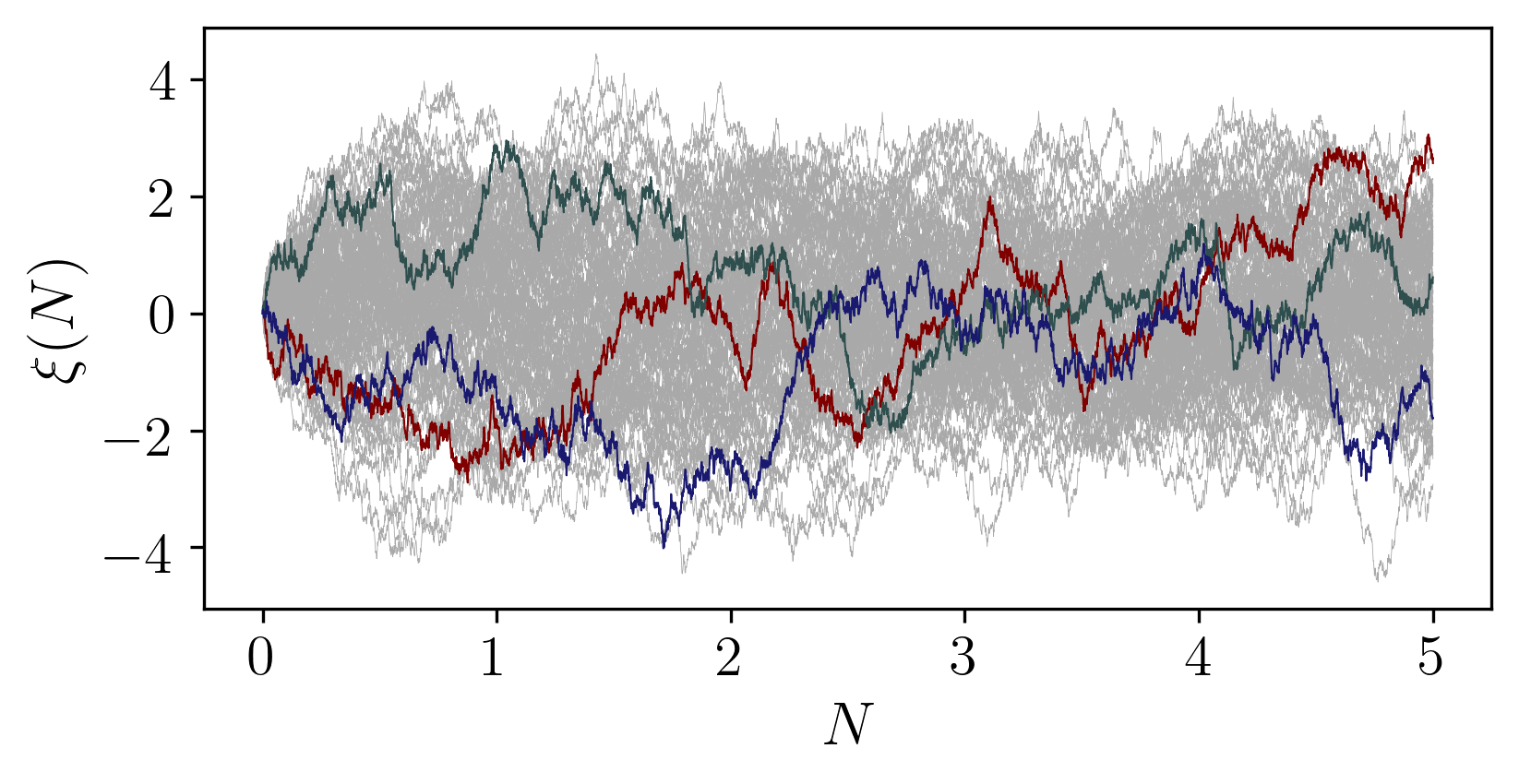}
\includegraphics{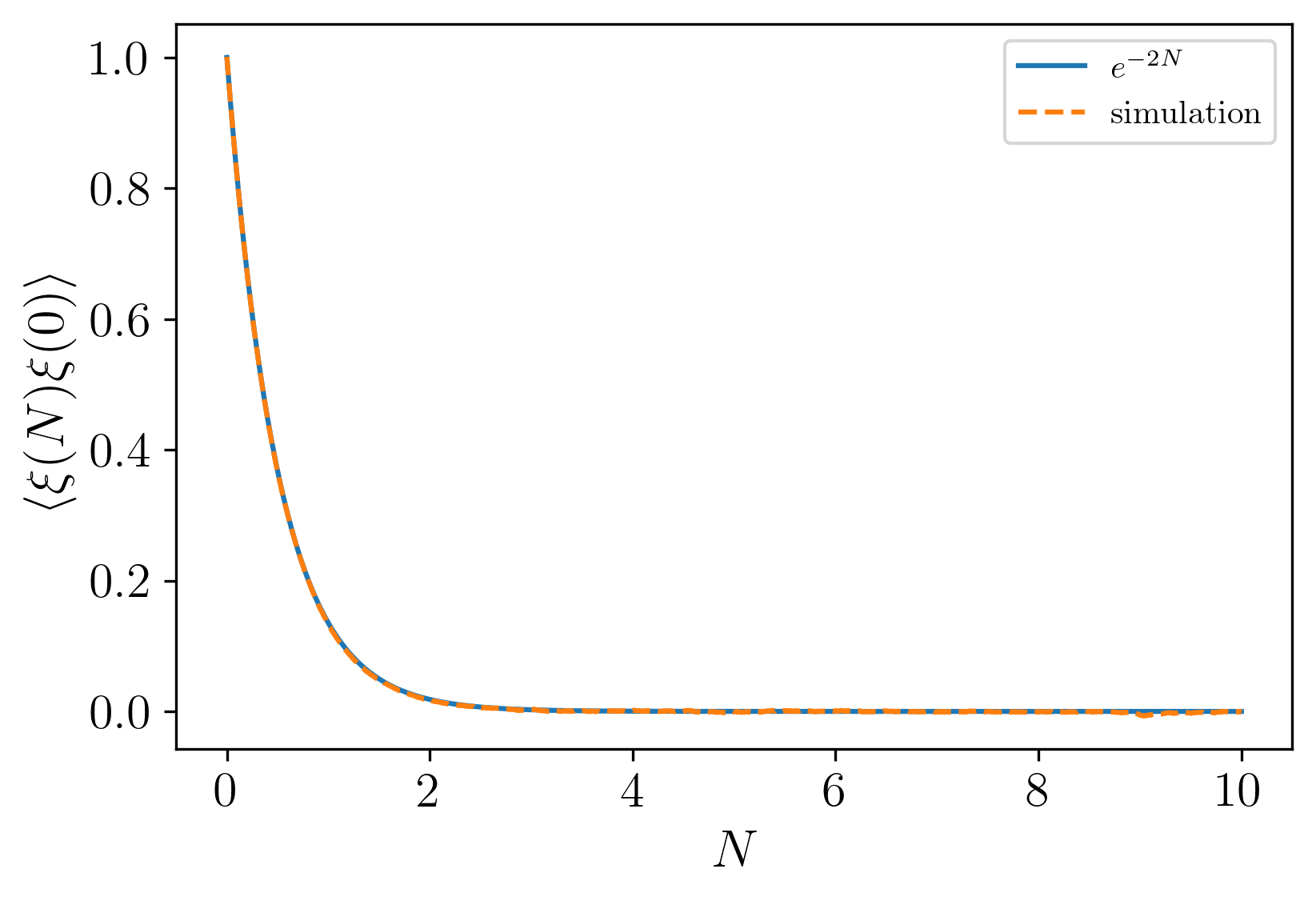}
\caption{Sample paths of the noise $\xi$ over several realizations with three of them colorized. The fact that they hover around the origin shows that $\big\langle \xi(N) \big\rangle=0$ (top panel). However, it can be shown that these paths have a decaying exponential correlation (bottom panel).}
\label{fig:process}
\end{figure}
$W(t)$ denotes a Wiener process. We have defined $\eta(t)\equiv\frac{\dd W}{\dd t}$ as the derivative of the Wiener process and $\xi(t)$ is the stochastic process that gives rise to the required correlated two-point function. This follows from the property of Wiener processes that $\dd W\simeq\sqrt{\dd t}\mathscr{N}(0,1)$, with $\mathscr{N}(0,1)$ being the normal distribution with zero mean and unit variance. No such procedure exists for a general stochastic process. As a result, generation of the required stochastic process $x$, which is sourced by $\xi(t)$, requires knowledge of $\xi(t)$ itself -- the latter being sourced by Gaussian white noise possessing the desired statistics. As previously mentioned, this is precisely why numerically solving an SDE driven by colored noise is difficult. Unlike the Wiener process, a general stochastic process cannot be discretized without knowledge of how it is distributed in consecutive time steps, \textit{i.e.}, how to evaluate $\xi(t_{i+1})-\xi(t_{i})$ in a probabilistic sense.\\
\indent We now provide a simple discretization procedure for the inflaton-Langevin equation and the OU noise. To keep this procedure simple, the Euler-Maruyama method \cite{gardiner1985handbook} is used. Then, we have
\begin{align}
\xi_{i+1}&=\xi_i -2\xi_i \Delta N+2\sqrt{2}\underbrace{\sqrt{\Delta N}\mathscr{N}(0,1)}_{\text{Wiener process}}\;\;\;\;\;\;\text{with }\xi_0\sim\mathscr{N}(0,1)\\
\bar{\phi}_{i+1}&=\bar{\phi_i}-\frac{\partial_{\phi}V(\bar{\phi}_i)}{V(\bar{\phi}_i)}\Delta N + \frac{H(\bar{\phi}_i)}{2\pi}\xi_i\Delta N
\end{align}
The sample paths of several realizations of the exponentially correlated noise is shown in the top panel of Fig. (\ref{fig:process}) where we also verify that the OU noise generates the required exponentially decaying noise correlation function, shown in the bottom panel. Correlated noise can also be generated using a more straightforward, `brute-force' approach by using a fast-Fourier-transform (FFT) \cite{PhysRevA.42.7492,PhysRevA.46.8028}. But such a method is useful when one has little or no information on how to derive the auxiliary SDE.

\subsection{Numerical tests with Ornstein-Uhlenbeck noise}
Here we numerically compute the power spectrum of curvature perturbations $\mathcal{P}_{\zeta}$ and average $e$-folds $\big\langle \mathcal{N} \big\rangle$ by solving the inflaton-Langevin equations using the LO noise correlation function for $n=2$. The curvature power spectrum obtained by considering colored noise can be compare to the ones obtained from white noise and standard Mukhanov-Sasaki results. In the stochastic inflation formalism, one way to compute the curvature perturbation is to exploit the first order variation of the inflaton field with respect to the background, $\big\langle \delta\phi^{(1)} \big\rangle$. The angular terms refer to stochastic averages, which are defined as follows \cite{Ezquiaga:2018gbw,De:2020hdo}
\begin{equation}
\Big\langle \delta\phi^{(1)\;n} \Big\rangle=\frac{1}{n_{\text{sim}}}\sum_{i=1}^{n_{\text{sim}}}\left( \bar{\phi}-\phi_{\text{bg}} \right)^{n}_{i}
\end{equation}
\begin{figure}
\centering
\includegraphics[scale=0.5]{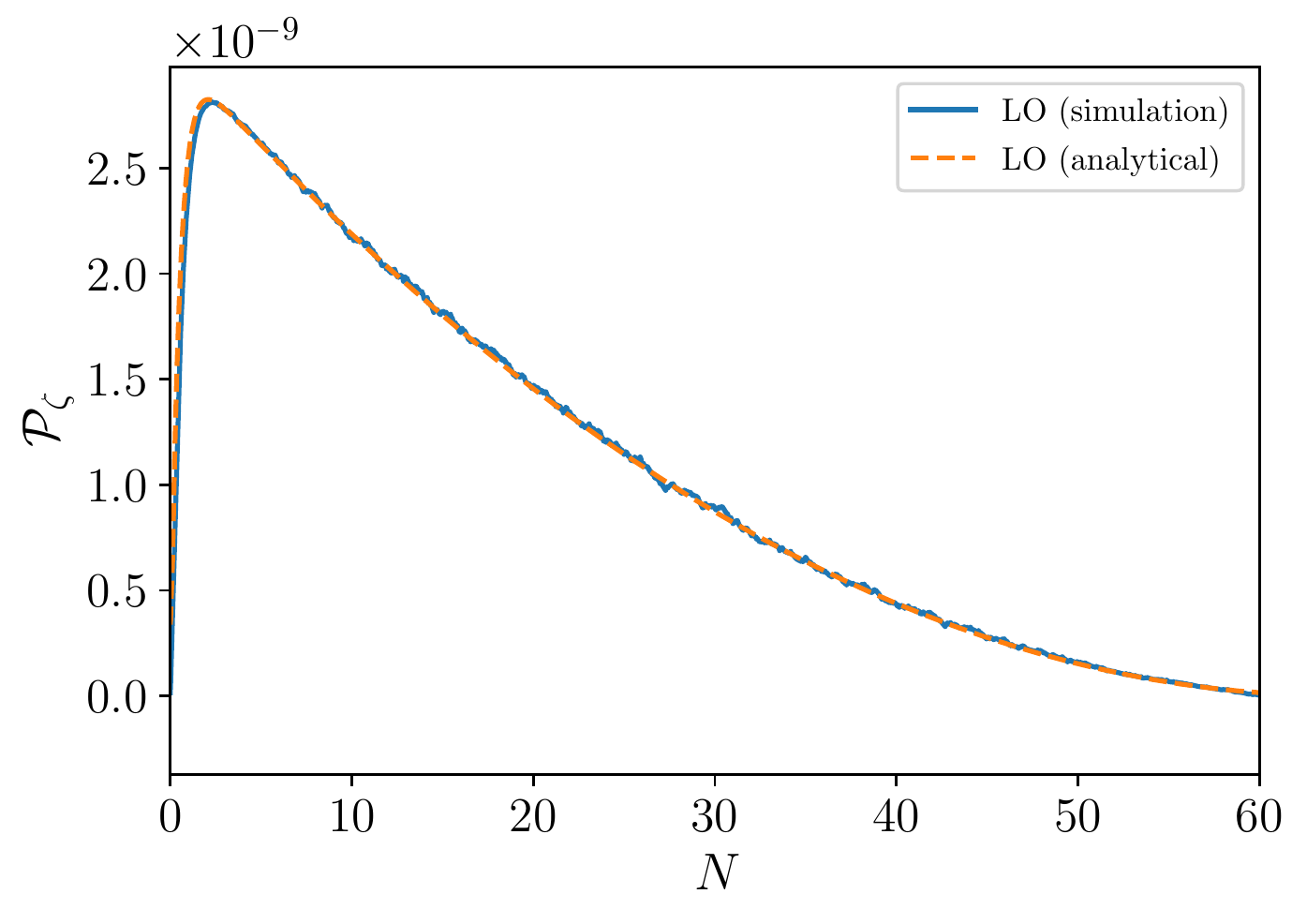}
\includegraphics[scale=0.5]{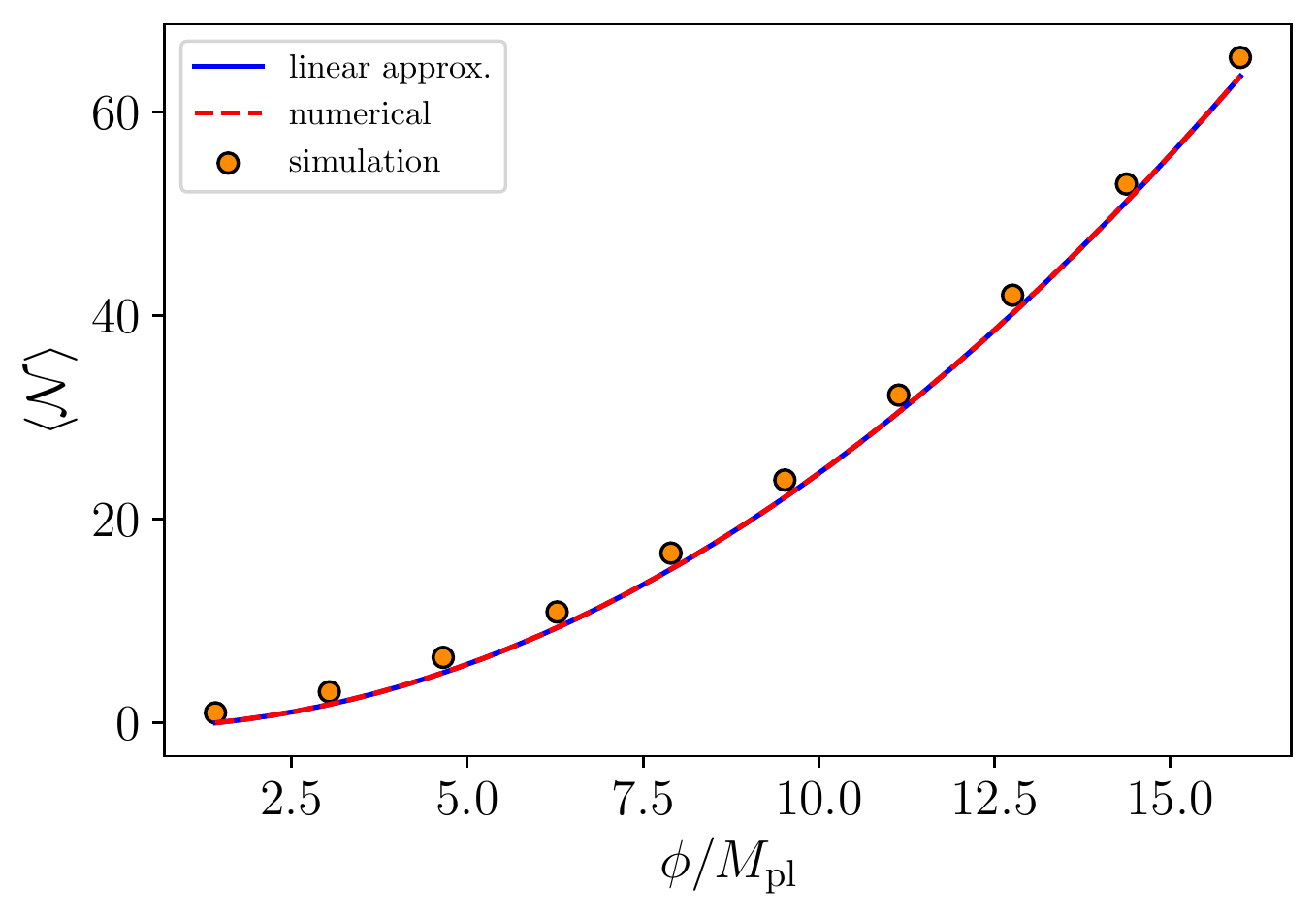}
\caption{Power spectrum of curvature perturbations $\mathcal{P}_\zeta$ (left panel) and average number of $e$-folds (right panel) for the LO noise correlation function with $n=2$. The results were obtained for $n_\text{sim}=10^6$.}
\label{fig:sim}
\end{figure}
where $\phi_\text{bg}$ refers to the classical background evolution. In principle, if the inflaton-Langevin equations are simulated for large $n_{\text{sim}}$, any number of these stochastic averages can be computed efficiently and no two trajectories will be the same due to the sourcing from the noise term. For the calculation of $\mathcal{P}_{\zeta}$ we use the definition of the power spectrum derived from the variance of the curvature perturbation \cite{Lyth:2009zz}
\begin{equation}
\Big\langle \zeta^{2}(\bm{x}) \Big\rangle=\int\dd\ln k\;\mathcal{P}_{\zeta}(k)\Longrightarrow\mathcal{P}_{\zeta}(k)=\frac{\dd}{\dd\ln k}\Big\langle \zeta^2 \Big\rangle
\end{equation}
Describing curvature perturbations in the spatially flat gauge amounts to $\zeta = \delta\phi/\sqrt{2\epsilon_{1}}$. We further change the variable $k$ to $e$-folds. Then, we have\footnote{It is worth stating some important issues concerning gauge choices. We define $\langle \delta\phi^{(1)\:2} \rangle$ in the spatially-flat gauge while the inflaton-Langevin equations are written in the uniform-$N$ gauge. Therefore, one needs to transform from the spatially-flat to the uniform-$N$ gauge before defining the noise correlation functions. However, in \cite{Pattison:2019hef} it was shown that corrections coming from such a gauge transformation is subdominant in slow-roll.}
\begin{equation}
\mathcal{P}_{\zeta}(N)=\frac{1}{1-\epsilon_{1}}\frac{\dd}{\dd N}\left( \frac{\Big\langle \delta\phi^{(1)\;2} \Big\rangle}{2\epsilon_{1}} \right)
\end{equation}
Similarly, $\big\langle \mathcal{N} \big\rangle$ is calculated by simulating the SDEs $n_\text{sim}$ times until the slow-roll condition is violated, with the procedure performed for a range of field values. The results of the numerical simulations are plotted in Fig. (\ref{fig:sim}). In the left panel, the simulated (solid blue) and analytical (dashed orange) $\mathcal{P}_\zeta$ are shown. In the right panel, the values of $\big\langle \mathcal{N} \big\rangle$ for ten different field values are included with the results shown in Fig. (\ref{fig:ave_efold}). The plots show that the colored noise dynamics can be captured effectively using the leading order expansion of the exact noise two-point correlation function. Furthermore, the small $e$-fold suppression of $\mathcal{P}_\zeta$ is well reproduced using the numerical scheme and will yield the similar results for different values of $n$. Nevertheless, certain small deviations are expected to exist between the analytical calculations and simulation results. For one, we know that the OU correlation is not exactly a decaying exponential but contains an additional $e^{(t_1+t_2)/\tau}$ piece, as seen in Eq. \eqref{eq:OU_correlation_full}. Although negligible at later times, this piece may, nevertheless, produce small, yet noticeable differences.

\section{Conclusion}

In this paper, the stochastic inflationary dynamics of single field inflation models were formulated where the coarse-grained dynamics is described by a general class of exponential window functions. Using the slow-roll and massless de Sitter approximations, we showed that the noise two-point correlation function behaves as $\big\langle \xi(N)\xi(N') \big\rangle_{(n)}\sim \left( \cosh\left[ n(N-N') \right]+1 \right)^{-1}$. Using this, we showed that the power spectrum of curvature perturbations $\mathcal{P}_\zeta$ experiences a period of suppression during small $e$-folds, before joining up with the expected slow-roll result at large $e$-folds. Exact expressions for $\mathcal{P}_\zeta$ were derived for the chaotic inflation model only (along with early and late $e$-fold asymptotics) while the Starobinsky model was also numerically studied. Moreover, a first passage time analysis was performed by constructing an approximate Fokker-Planck equation. This was, of course, derived under the assumption that the noise two-point correlation is exponentially decaying (Ornstein-Uhlenbeck). It was shown that the analytical results for $\big\langle \mathcal{N} \big\rangle$ and $\mathcal{P}_\zeta$ were backed up by numerical simulations of the inflaton-Langevin equations, especially the fact that early $e$-fold suppression of $\mathcal{P}_\zeta$ is also visible from numerical calculations.\\
\indent In the matter of choosing the proper auxiliary SDE to source the inflaton evolution, we advocate the use of the leading order term in the expansion of the noise two-point correlation function, which is nothing other than the Ornstein-Uhlenbeck correlation function. It is relatively simple to implement while also being able to reproduce the effects of the full correlation function to a good degree of accuracy. If one wishes to simulate the stochastic processes in complete generality, then the derivation of a new auxiliary SDE is required such that it possesses the required correlation function, which is demonstrably a nontrivial task. The alternative, as mentioned previously, is to generate the noise using FFT using a known power spectral density. We have not performed calculations using the FFT method, so it is not known how it will perform compared to the auxiliary SDE method. This can be explored in a future work. However, since the FFT method requires a power spectral density, the situation can be made more complicated if one is unable to obtain (semi)-analytical expressions for it. Fortunately for our case, the power spectral density can be calculated analytically since the since the same can be done to the Fourier transform of the noise correlation function. Moreover, we mention again that we have used the It\^{o} scheme for discretizing the noise (corresponding to $\alpha=0$). If the noise were additive, no difference would exist among the many choices. However, since $H(\bar{\phi})$, the noise is multiplicative and, thus, discretization dependent. One can, of course, perform the simulations using Stratonovich, which will amount to an additional noise-induced drift term. However, since $H(\bar{\phi})$ is only slowly varying for a large part of slow-roll, there should not be appreciable differences between the two.\\
\indent It was discussed in Refs. \cite{Linde:2001ae,Piao:2005ag,Destri:2009hn} that a period of kinetic domination (fast roll) preceding the inflationary stage can help provide a better fit to the CMB power spectrum at low-$\ell$ multipoles. One of the characteristics of fast roll is a suppression of $\mathcal{P}_\zeta$ \cite{Ragavendra:2020old}, which is rather intuitive since ultra slow-roll produces the opposite effect. Similarly, smooth coarse-graining induces a suppression of the curvature power spectrum and one can certainly study its effects on the CMB power spectrum. This is a distinctly different mechanism, though, since fast-roll inflation relies on a portion of the potential that is steep while in our case the suppression is purely a consequence of the finite correlation time of the noise. In order to do that, one should opt for a small value of $n$ in a way that the suppression lasts for an appreciable number of $e$-folds, after which the CMB power spectrum can be calculated using a numerical suite such as \textsf{CAMB} \cite{camb}, although one should be mindful of the fact that relating the Fourier modes $k$ to the $e$-folds $N$ in stochastic inflation is not as simple as $k=a(N)H(N)$, as shown in Ref. \cite{Ando:2020fjm} since there might not be a simple, one-to-one correspondence between $k$ and $\phi$. However, the authors showed that the standard mapping between $k$ and $N$ is valid in the low-diffusion regime (when quantum diffusion always plays a subdominant role compared to classical drift).\\
\indent The main conclusion from our work is that in the late $e$-fold times, there is essentially negligible difference in the power spectrum based on the kind of window function one uses. The difference only arises at small $e$-folds. Here we clarify that the use of `early' and `late' to describe different regimes are made in regards to the $e$-fold variable $N$ that is used when solving the inflaton-Langevin equations. The regime of applicability of the stochastic inflation formalism is that of large $e$-folds and according to the present work, it is not very sensitive to the kind of window function one uses.

\section{Acknowledgements}
The authors thank Joseph Kapusta and Swagat Mishra for helpful comments regarding the manuscript. RM is supported by the University of Minnesota Doctoral Dissertation Fellowship. AD acknowledges the support of U.S. DOE Grant No. DEFG02-87ER40328. We also thank the anonymous referee for suggesting helpful changes and corrections over the span of the review.
\bibliographystyle{JHEP}
\bibliography{refs}

\appendix
\section{Asymptotic expressions for $W_{(n)}(s)$ for even $n$}\label{sec:asymptotics}
\begin{figure}
    \centering
    \begin{tikzpicture}
        \fill[blue!25] (-3.5,-0.1) rectangle (3.5,0.1);
        \draw[thick,->] (-5,0) -- (5,0)node[right]{$\Re{f(z)}$};
        \draw[thick,->] (0,-1) -- (0,3)node[above]{$\Im{f(z)}$};
        \draw[thick,->] (2,1.5) -- (3,0.5)node[right]{\tiny $\mathcal{C}_1:z=z_{0,n}+re^{i\theta_1}$};
        \draw[thick,->] (-3,0.5)node[left]{\tiny $\mathcal{C}_2:z=z_{n/2-1,n}+re^{i\theta_2}$} -- (-2,1.5);
        \fill[red] (2.5,1) circle (2pt)node[right]{$z_{0,n}$};
        \fill[red] (-2.5,1) circle (2pt)node[left]{$z_{n/2-1,n}$};
        \node[scale=1.0] at (-3.5,-0.3){to $-\infty$};
        \node[scale=1.0] at (3.5,-0.3){to $+\infty$};
    \end{tikzpicture}
    \caption{Illustration of integration contours through steepest descent paths.}
    \label{fig:contours}
\end{figure}
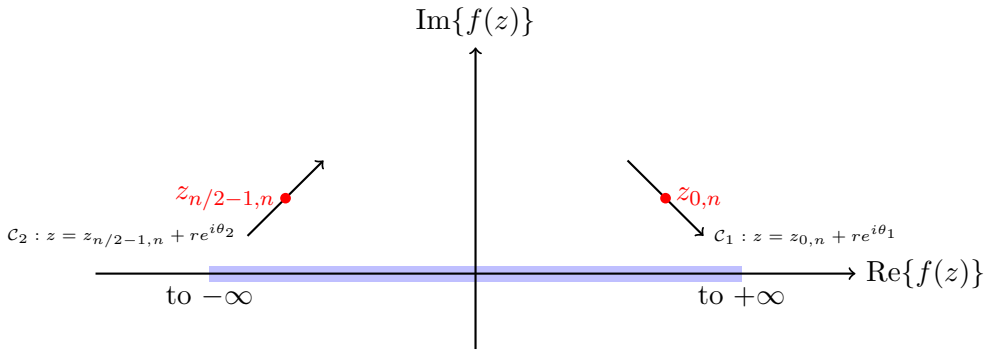

In order to check whether Eq. \eqref{eq:condition} is satisfied for the real space window function for arbitrary $n$, we can performed a generalized asymptotic calculation of $W_{(n)}(s)$ as $s\rightarrow\infty$. The real space window function can be written as
\begin{equation}
    R_\sigma^3 W_{(n)}(s)=\frac{1}{2\pi^2 s}\int_{0}^{\infty}\dd u\:\sin\left( su \right)ue^{-u^n /2}
\end{equation}
where $u=kR_\sigma$ and $s=r/R_\sigma$. To solve this integral in the asymptotic limit, we first transform it to the complex plane and make use of the method of steepest descent. Then the integral becomes
\begin{equation}
    \frac{1}{2}\Im{\int_{-\infty}^{\infty}\dd u \:u \exp\left[ -\frac{1}{2}\left( u^n-2isu \right) \right]}
\end{equation}
Furthermore, with the redefinition $u=zs^{\frac{1}{n-1}}=\alpha_{n,s}z=\frac{\beta_{n,s}}{s}z$, we have
\begin{equation}
    \frac{1}{2}\alpha^2_{n,s}\Im{\int_{-\infty}^{\infty}\dd z\:z \exp\left[ -\frac{\beta_{n,s}}{2}\left( z^n-2iz \right) \right]}=\frac{1}{2}\alpha^2_{n,s}\Im{\mathcal{I}(s)}
\end{equation}
To apply the method of steepest descent, we first compute the saddle points of $f(z)=z^n-2iz$. They are
\begin{equation}
    z_{k,n}=\left( \frac{2}{n} \right)^{\frac{1}{n-1}}\exp\left[ \frac{i\pi(1+4k)}{2(n-1)} \right]\;\;\;\;\;\;(k=0,1,\cdot\cdot\cdot, n-2)
\end{equation}
We can argue that, for even $n$, there are two symmetric saddle points that provide the greatest contributions to $\mathcal{I}(s)$. These are $z_{0,n}$ and $z_{n/2-1,n}$. We define contours $\mathcal{C}_1$ and $\mathcal{C}_2$ passing through $z_{0,n}$ and $z_{n/2-1,n}$ respectively, such that
\begin{equation}
    \mathcal{I}(s)=\mathcal{I}^{(1)}(s)+\mathcal{I}^{(2)}(s)
\end{equation}
Moreover, the steepest descent paths passing through the saddle points in the correct direction are $\theta_1=-\frac{\pi}{2(n-1)}$ and $\theta_2=\frac{\pi}{2(n-1)}$ (see Fig. (\ref{fig:contours})). Considering $\mathcal{C}_1$ and the steepest descent path $z-z_{0,n}=re^{-\frac{i\pi}{2(n-1)}}$, we can write
\begin{equation}
    \mathcal{I}^{(1)}=e^{-\frac{i\pi}{2(n-1)}}\int_{-\infty}^{\infty}\dd r\left( z_{n,0}+re^{-\frac{i\pi}{2(n-1)}} \right)\exp\left[ -\frac{\beta_{n,s}}{2}f(z_{0,n}) \right]\exp\left[ -\frac{\beta_{n,s}}{4}f''(z_{0,n})r^2 e^{-\frac{i\pi}{n-1}} \right]
\end{equation}
It turns out that $\mathcal{I}^{(2)}=-[\mathcal{I}^{(1)}]^*$. Carrying out the saddle point integration, we find 
\begin{align}
    R^3_\sigma W_{(n)}(s)\sim\frac{\alpha^2_{n,s}}{4\pi s}\left( \frac{2}{n} \right)^{\frac{1}{n-1}}e^{-\frac{\beta_{n,s}}{2}\Re{f(z_{0,n})}}\sqrt{\frac{4\pi}{\beta_{n,s}|f''(z_{0,n})|}}\sin\left[ \frac{\beta_{n,s}}{2}\Im{f(z_{0,n})}-\arg{z_{0,n}} \right]
\end{align}
Substituting the expressions for $\alpha_{n,s}$ and $\beta_{n,s}$, we have
\begin{equation}\label{eq:asymptotic_general}
    R^3_\sigma W_{n}(s)\sim \frac{1}{4\pi}\sqrt{\frac{\pi}{3}}s^{-\frac{3}{2}\frac{n-2}{n-1}}e^{-\frac{1}{2}\Re{f(z_{0,n})}s^{\frac{n}{n-1}}}\sin\left[ \frac{s^{\frac{n}{n-1}}}{2}\Im{f(z_{0,n})}-\arg{z_{0,n}} \right]
\end{equation}

\section{Small $e$-fold behavior of $\mathcal{P}_{\zeta}$ with colored noise --- a back-of-the-envelope calculation}\label{small_efold}
We will generalize a behavior of stochastic averages for the case of colored noise in the small $e$-fold limit. The quantity $\big\langle \xi\delta\phi^{(1)} \big\rangle$ is constant in the white noise case. However, with the $\left( \cosh\left[ n(N_1-N_2) \right]+1 \right)^{-1}$ noise, it can be shown that this quantity displays a universal behavior. Here, we derive the characteristic variations of these averages through a series of approximations generally applicable for small $e$-fold intervals. First, we focus on the exponential term in the integral of Eq. \eqref{eq:expec}
\begin{equation}
\exp\left[ 2M_{\text{pl}}^{2}\int^{N_{1}}\dd N_{2}\;\left( \frac{H'}{H} \right)' \right]
\end{equation}
While the inflaton is in the slow-roll attractor, this exponential changes very slowly. Since $H^{2}\simeq V/3M^{2}_{\text{pl}}$, we have $H'/H\approx\sqrt{2\epsilon_{\text{\tiny V}}}/2M_{\text{pl}} $, where $\epsilon_{\text{\tiny V}}=M_{\text{pl}}^2(V'/V)^2 /2$ is the first slow-roll parameter \cite{baumann2012tasi}. With another derivative with respect to the field variable and we get
\begin{equation}
\int\dd N_{2}\;\left( \frac{H'}{H} \right)'\approx \int\dd N_{2} \left( \eta_{\text{\tiny V}}-2\epsilon_{\text{\tiny V}} \right)
\end{equation}
with $\eta_{\text{\tiny V}}=M_{\text{pl}}^{2}V''/V$ being the second slow-roll parameter. During the start of inflation, the slow-roll parameters are very small -- $\epsilon_{\text{\tiny V}},\eta_{\text{\tiny V}}\sim\mathcal{O}(10^{-2})$ for quadratic potentials -- and vary slowly with time. If we consider a very small $e$-fold interval $\Delta n$ at the start of inflation, $e^{(2\eta_{\text{\tiny V}}-4\epsilon_{\text{\tiny V}})\Delta n}\simeq e^{(n_{s}-1)\Delta n}\sim 1$.\footnote{We have used definition of the scalar spectral index $n_s$ in terms of slow-roll parameters. It is usually expressed as $$ n_s\simeq1-4\epsilon+2\eta $$Although $\epsilon$ and $\eta$ are defined in terms of the Hubble parameter, the variants defined in terms of $V(\phi)$ are valid in slow-roll.} Hence, for small $e$-folds, we have the following behavior for Eq. \eqref{eq:expec}
\begin{equation}
\Big\langle \xi\delta\phi^{(1)} \Big\rangle_{(n)}\simeq\frac{H}{4\pi}\tanh\left(\frac{n}{2}N\right)
\end{equation}
It is clear the $\big\langle \xi\delta\phi^{(1)} \big\rangle$ behaves as if `switched on' due to the presence of the $\tanh$ function before joining up with the white noise (or classical) result, with the $e$-fold time of this switching dependent on the value of $n$. In the same manner, we can further show that such a feature is carried on to the curvature power spectrum such that there is a small suppression of $\mathcal{P}_{\zeta}$ at large scales (large scales being synonymous to the large length scale fluctuations that usually become superhorizon during the early stages on inflation). Still working in the slow-roll approximation, the variance of the field perturbations simplifies to
\begin{equation}
\Big\langle \delta\phi^{(1)\;2} \Big\rangle_{(n)}\sim\frac{H^2}{4\pi^2}\frac{2}{n}\ln\left[ \cosh\left( \frac{n}{2}N \right) \right]
\end{equation}
with the power spectrum given reducing to
\begin{align}\label{eq:PzetaApprox_colored}
\mathcal{P}_{\zeta}&= \frac{\dd}{\dd N}\left( \frac{\Big\langle \delta\phi^{(1)\;2} \Big\rangle}{2M_{\text{pl}}^2\epsilon_{1}} \right)\nonumber\\
&\simeq\frac{1}{2M_{\text{pl}}^2\epsilon_{1}}\frac{\dd}{\dd N}\Big\langle \delta\phi^{(1)\;2} \Big\rangle
\end{align}
\begin{equation}\label{eq:Pzeta_asymp1}
\boxed{\mathcal{P}_{\zeta}^{0\lesssim N\lesssim \Delta n}\simeq\frac{H^2}{8\pi^{2}M_{\text{pl}}^2\epsilon_{1}}\tanh \left(\frac{n}{2}N\right)\simeq\frac{H^2}{8\pi^{2}M_{\text{pl}}^2\epsilon_{1}}\left[ \frac{n}{2}N-\frac{1}{3}\left(\frac{n}{2}N\right)^3 \right]}
\end{equation}

Again, the power spectrum is almost similar to the slow-roll expression apart from the small $e$-fold modification arising from the colored noise sourcing of the inflaton-Langevin equation. The suppression of the power spectrum in Eq. \eqref{eq:PzetaApprox_colored} is surprising. Furthermore, since the dynamics is in slow-roll, one can alternatively express $\mathcal{P}_\zeta$ in the familiar power law form. Depending on the value of $n$ in the window function, one can define a cut-off Fourier mode $k_c$ up to which the power spectrum suppression lasts. Hence, $\mathcal{P}_\zeta$ can be modeled as a broken power law of the following form

\[
\mathcal{P}_\zeta (k)=
\begin{dcases}
k^\alpha & (k<k_c)\\
\mathcal{P}(k_\star)\left( \frac{k}{k_\star} \right)^{n_s -1} & (k>k_c)
\end{dcases}
\]
where $\alpha>0$ can be found through interpolation.

\section{Formulating the first passage time analysis}\label{sec:first_passage_deriv}
We consider inflation taking place in the range of field values given by $\phi_\text{end}\leq\phi\leq\phi_\text{uv}$. We have already seen that the FPE describes the evolution of the conditional probability distribution $P(\phi,N|\phi_\text{in},N_\text{in})$, in the sense that it evolves it forward in time. In the formulation of the first passage time problem, we first derive the adjoint FPE (backward Kolmogorov equation). Using the Chapman-Kolmogorov equation, the conditional probability can be expressed in terms of some intermediate state
\begin{equation}
P(\phi,N|\phi_\text{in},N_\text{in})=\int\dd\phi'\:P(\phi,N|\phi',N')P(\phi',N'|\phi_\text{in},N_\text{in})
\end{equation}
It states that, in going from $(\phi_\text{in},N_\text{in})$ to $(\phi, N)$, the inflaton can go through any number of intermediate states $(\phi',N')$. Differentiating with respect to this intermediate point, we get
\begin{equation}
0=\int\dd\phi'\left[ \left(\frac{\partial}{\partial N'}P(\phi,N|\phi',N')\right)P(\phi',N'|\phi_\text{in},N_\text{in})+P(\phi,N|\phi',N')\mathcal{L}_\text{FP}(\phi')\cdot P(\phi',N,|\phi_\text{in},N_\text{in}) \right]
\end{equation}
Using the second term in the integral and integrating it by parts, the adjoint FPE operator can be defined
\begin{align}
\int_{\phi_\text{uv}}^{\phi_\text{in}}\dd\phi'P(\phi,N|\phi',N')\mathcal{L}_\text{FP}(\phi')\cdot P(\phi',N,|\phi_\text{in},N_\text{in})&=\int_{\phi_\text{uv}}^{\phi_\text{in}}\dd\phi'\left[\mathcal{L}^{\dagger}_{\text{FP}}(\phi')\cdot P(\phi,N|\phi',N')\right]\nonumber\\
&\times P(\phi',N'|\phi_\text{in},N_\text{in})+\text{vanishing surface terms}
\end{align}
where
\begin{align}
-\frac{\partial}{\partial N_\text{in}}P(\phi,N|\phi_\text{in},N_\text{in})&=\left[ \mu(\phi_\text{in})\frac{\partial}{\partial\phi_\text{in}}+\frac{1}{2}\widetilde{\sigma}^2(\phi_\text{in})\frac{\partial^2}{\partial\phi^2_\text{in}} \right]P(\phi,N|\phi_\text{in},N_\text{in})\nonumber\\
&=\mathcal{L}^{\dagger}_\text{FP}(\phi_\text{in})\cdot P(\phi,N|\phi_\text{in},N_\text{in})
\end{align}
Although the forward and backward Kolmogorov equations are equivalent, the latter is useful in first passage problems where we are interested in calculating how long it takes for a particle to exit a certain region (in our case, how long inflation lasts). This is ensured by placing an absorbing barrier at $\phi_\text{end}$ such that $P(\phi_\text{end},N|\phi_\text{in},N_\text{in})=0$. To that end, we compute the survival probability $\mathcal{S}(N)$ that the inflaton remains in the interval $\phi_\text{uv}\leq\phi\leq\phi_\text{end}$ at time $N$
\begin{equation}
\mathcal{S}(N)=\int_{\phi_\text{uv}}^{\phi_\text{end}}\dd\phi\: P(\phi,N|\phi_\text{in},N_\text{in})
\end{equation}
This also corresponds to the probability that the inflaton will have left the interval at $\mathcal{N}>N$. Then, we also have
\begin{equation}
\mathcal{S}(N)=\int_N^\infty \dd N'\:P(N')
\end{equation}
or,
\begin{equation}\label{eq:prob}
P(N)=-\frac{\dd \mathcal{S}(N)}{\dd N}=-\int_{\phi_\text{uv}}^{\phi_\text{end}}\dd\phi\: \frac{\partial}{\partial N}P(\phi,N|\phi_\text{in},N_\text{in})
\end{equation}
This can be equivalently rephrased as follows -- since $\mathcal{S}(N)$ is the probability that the inflaton is within the defined boundaries, the probability it has escaped is $1-\mathcal{S}(N)$, hence the rate at which it escapes is given by the $e$-fold derivative. Using Eq. \eqref{eq:prob}, the $n$-th moments of the $e$-fold distribution can be calculated.
\begin{align}
\Big\langle \mathcal{N}^n \Big\rangle(\phi_\text{in})&=\int_{N_\text{in}}^\infty \dd N\:N^n P(N)\nonumber\\
&=-\int_{N_\text{in}}^\infty \dd N\:N^n\int_{\phi_\text{uv}}^{\phi_\text{end}}\dd\phi\: \frac{\partial}{\partial N}P(\phi,N|\phi_\text{in},N_\text{in})
\end{align}
Using integration by parts, we arrive at
\begin{equation}\label{eq:N_average}
\Big\langle \mathcal{N}^n \Big\rangle=n\int_{N_\text{in}}^\infty \dd N\:N^{n-1}\int_{\phi_\text{uv}}^{\phi_\text{end}}\dd\phi\: P(\phi,N|\phi_\text{in},N_\text{in})-\int_{\phi_\text{end}}^{\phi_\text{uv}}\dd\phi\left[ P(\phi,\infty|\phi_\text{in},N_\text{in})-\underbrace{P(\phi,N_\text{in}|\phi_\text{in},N_\text{in})}_{\delta(\phi-\phi_\text{in})} \right]
\end{equation}
We require that $ P(\phi,\infty|\phi_\text{in},N_\text{in})$ vanish while the other term is a constant. Applying the adjoint Fokker-Planck operator to Eq. \eqref{eq:N_average} and employing the Markov property where the system depends on $N-N_\text{in}$, we have
\begin{align}
\mathcal{L}^{\dagger}_\text{FP}(\phi_\text{in})\cdot\Big\langle \mathcal{N}^n \Big\rangle(\phi_\text{in})&=-n\int_{N_\text{in}}^\infty \dd N\:N^{n-1}P(N)\\
&=-n\Big\langle \mathcal{N}^{n-1} \Big\rangle(\phi_\text{in})
\end{align}
At the end of the computation, we see that the statistics of the $e$-folds follow a hierarchy, connected together by the adjoint FPE.

\section{Derivation of the Ornstein-Uhlenbeck correlation function}\label{eq:OU_process_derivation}
We consider an OU process described by the following SDE
\begin{equation}\label{eq:OU_process}
\frac{\dd\xi(t)}{\dd t}=-\frac{1}{\tau}\xi(t)+\frac{1}{\tau}\eta(t)
\end{equation}
where $\eta(t)$ is a Gaussian white noise term with two-point correlation function $\big\langle \eta(t)\eta(t') \big\rangle=N\delta(t-t')$. Not only can Eq. \eqref{eq:OU_process} be analytically solved, it can also be discretized in a straightforward fashion using the properties of Wiener process. To solve Eq. \eqref{eq:OU_process}, we multiply both sides by $e^{t/\tau}$ and integrate. Then,

\begin{align}
\int_{0}^{t}\frac{\dd}{\dd t'}\left( e^{t'/\tau}\xi(t') \right)\dd t'&=\int_{0}^{t}\frac{e^{t'/\tau}}{\tau}\eta(t')\dd t'\nonumber\\
\xi(t)&=\xi(0)e^{-t/\tau}+\int_{0}^{t}\frac{e^{-(t-t')/\tau}}{\tau}\eta(t')\dd t' \label{eq:xi_sol}
\end{align}
Now, computing the two-point correlation function of Eq. \eqref{eq:xi_sol}, we get
\begin{align}\label{eq:OU_correlation_full}
\Big\langle \xi(t_1)\xi(t_2) \Big\rangle&=\frac{1}{\tau^2}\Bigg\langle \int_0^{t_1}\dd t'\:e^{-(t_1-t')/\tau}\eta(t')\int_0^{t_2}\dd t''\:e^{-(t_2-t'')/\tau}\eta(t'') \Bigg\rangle\nonumber\\
&=\frac{1}{\tau^2}\int_0^{t_1}\dd t'\int_0^{t_2}\dd t''e^{-(t_1+t_2-t'-t'')/\tau}\underbrace{\Big\langle \eta(t')\eta(t'') \Big\rangle}_{N\delta(t'-t'')}\nonumber\\
&=\frac{N}{\tau^2}\int_0^{\text{min}(t_1,t_2)}\dd t'\:e^{-(t_1+t_2-2t')/\tau}\nonumber\\
&=\frac{N}{2\tau}\left( e^{-|t_1-t_2|/\tau}-e^{-(t_1+t_2)/\tau} \right)
\end{align}
The second term in the parenthesis can be ignored and we see that the noise two-point correlation function for an OU process is a decaying exponential. For our case, with the leading-order noise term, we require $\big\langle \xi(N)\xi(N') \big\rangle\simeq 2e^{-2(N-N')}$. This can be achieved by setting $N=\tau^{-1}=2$. Equation \eqref{eq:OU_process} can be re-written for normalized noise correlation function as follows
\begin{equation}\label{eq:OU_process_final}
\frac{\dd\xi(N)}{\dd N}=-2\xi(N)+2\sqrt{2}\:\eta(N)
\end{equation}
Equation \eqref{eq:OU_process_final} can be discretized in a very straightforward manner by using the properties of the Wiener process-
\begin{equation}
\xi_{i+1}=\xi_i-2\xi_i\Delta N+2\sqrt{2}\left( W_{i+1}-W_{i} \right)
\end{equation}
where we have used the fact that $\eta(N)\dd N=\dd W(N)$. It can further be simplified by using the fact that increments of a Wiener process are independent and normally distributed, \textit{i.e.}, $W_{i+1}-W_i\simeq \mathcal{N}(0,\sqrt{\Delta N})$.

\end{document}